\documentclass[aps,prl,reprint]{revtex4-1}
\usepackage{amsmath}

\usepackage{color}
\usepackage{graphicx}% Include figure files
\usepackage{dcolumn}% Align table columns on decimal point
\usepackage{bm}% bold math
\usepackage{bbm}% bold math
\usepackage{setspace}
\usepackage{ulem}
\usepackage{braket}
\usepackage{multirow}
\usepackage{enumitem}

\newcommand{\til}{\tilde}
\newcommand{\ovl}{\overline}
\newcommand{\al}{\alpha}
\newcommand{\be}{\beta}

\newcommand{\si}{\sigma}
\newcommand{\ta}{\tau}
\newcommand{\ga}{\gamma}

\newcommand{\pa}{\partial}
\newcommand{\ka}{\kappa}
\newcommand{\de}{\delta}

\newcommand{\la}{\lambda}

\newcommand{\lan}{\langle}
\newcommand{\ran}{\rangle}

\newcommand{\vphi}{\varphi}

\newcommand{\Ltxx}{\text{\tiny L}}
\newcommand{\Rtxx}{\text{\tiny R}}

\newcommand{\Ctx}{\text{C}}

\newcommand{\dash}{\text{-1}}
\begin{document}

\title{Matrix product state formulation of the multiconfiguration time-dependent Hartree theory}

%\date{\today}
\date{November 8, 2018}

\author{Yuki Kurashige\footnote{Electronic mail: kura@kuchem.kyoto-u.ac.jp.}}
\affiliation{Department of Chemistry, Graduate School of Science, Kyoto University, Kitashirakawa Oiwake-cho, Sakyo-ku Kyoto, 606-8502, Japan}

\begin{abstract}
A matrix product state formulation of the multiconfiguration
time-dependent Hartree (MPS-MCTDH) theory is presented.
The Hilbert space that is spanned by the direct products of the phonon
degree of freedoms, which is linearly parameterized in the MCTDH
ansatz and thus results in an exponential increase of the
computational cost, is parametrized by the MPS form.
Equations of motion based on the Dirac-Frenkel time-dependent
variational principle is derived by using the tangent space projection
and the projector-splitting technique for the MPS, which have been
recently developed.
The mean-field operators, which appear in the equation of motion of
the MCTDH single particle functions (SPF), are written in terms of the
MPS form and efficiently evaluated by a sweep algorithm that is
similar to the DMRG sweep.
The efficiency and convergence of the MPS approximation to the MCTDH are
demonstrated by quantum dynamics simulations of extended excitonic
molecular systems.
\end{abstract}

\maketitle
%\setstretch{1.1}

\section{Introduction}
Matrix product state (MPS) is one of the most successful
tensor-network states (TNS) which encode quantum states in the
exponentially growing Hilbert space of strongly correlated systems to
a sequential product of tensors
and was first introduced by White \cite{White:1992ie, White:1993tn} as
an eigensolver named as the density-matrix renormalized group (DMRG)
algorithm.
Although the DMRG works best for one-dimensional Hamiltonians, it has
been successfully extended to many fields of application such as {\it
  ab initio} Hamiltonian of quantum chemistry where all the degree of
freedoms (DOFs), i.e.\ electrons, are coupled complexly through the
Coulomb interaction \cite{White:1999ws, Chan:2002cz}.
Extension to time-dependent simulations, i.e.\ time evolution of the
MPS, has been also developed, such as the adaptive time-dependent DMRG
($t$-DMRG) \cite{White:2004fd, Daley:2004hk} and the time-evolving
block-decimation (TEBD) algorithm \cite{Vidal:2004jc}.
The use of the Suzuki-Trotter decomposition by splitting the summation
of terms of interactions in the Hamiltonian makes the methods
particularly efficient, but in principle it works well for
Hamiltonians consisting of local interactions, typically nearest
neighbor interaction models.

Recently, time-evolution of the MPS wavefunction has been formulated
within the Dirac-Frenkel time-dependent variational principle (TDVP)
\cite{Haegeman:2011dz, Lubich:2013ku, Haegeman:2013bg}. This method,
in principle, is not restricted to any particular type of Hamiltonians
and reasonably applicable to long-range interactions including
two-dimensional systems. Efficiency of the method was, however,
deteriorated by numerical instability problems arose from the highly
nonlinear parametrization in the wavefunction ansatz.
It should be noted that there are other closely-related works,
e.g.\ Ref[\onlinecite{Ueda:2006tb, Dorando:2009dz, Kinder:2011tq,
    kinderquantum, wouters2013thouless, nakatani2014linear}].
Very recently, Haegeman and co-workers have developed a novel method
that circumvents those problems \cite{Haegeman:2016fo}. The method
utilizes Lie-Trotter splitting of the tangent space projectors of
different sites canonical form, which realizes an efficient and stable
propagation of the MPS wavefunctions. The development of robust time
evolution method based on the TDVP should significantly expand the
applicability of the MPS to a variety of
problems.\cite{Schroder:2016jza, Borrelli:2017jf, Kloss:2018be}

Time evolution methods based on the TDVP have great affinity for
molecular quantum dynamics simulations, in which the interactions in
the Hamiltonian are represented in the first quantization form.
The basis function expansion of the wavefunctions allows us to
efficiently evaluate the Hamiltonian matrix elements by analytical or
numerical integrations in the real space.
The multiconfiguration time-dependent Hartree (MCTDH) theory
\cite{Meyer:1990if, Manthe:1992cq, Beck:2000wm} is the most widely
used method in this field. In the method, variational space of a
vibrational wavefunction is spanned by direct products of one-particle
basis functions,
of which the shapes are also regarded as variational parameters and
evolved with time. While the use of variable one-particle basis
significantly reduces the size of the Hilbert space and makes the
method highly efficient,\cite{Worth:1996ck,Raab:1999fa} the number of
linear parameters of the space still grows exponentially with the
number of phonon modes. It is, therefore, natural to introduce TNS
ansatzes instead of the linear parametrization, in fact the
multi-layer formulation (ML-MCTDH), which corresponds to the
hierarchical Tucker tensor decomposition \cite{Grasedyck:2010cs}
and can avoid the expensive exponential cost,\cite{Manthe:2008ev,
  Wang:2015ef}
%
%}
%
made a great success\cite{Wang:2003fu, Manthe:2008ev,
  Vendrell:2011fh}
in many fields of application,\cite{Wang:2007eu, Cao:2013kx,
  Manthe:2017hj, Manthe:2017dx}
%
%}
%
in particular for encoding a tremendous
number of harmonic oscillators describing the bath modes that couple
to one of the molecules in molecular aggregates.\cite{Kuhn:2011tv,
  Schroter:2015je, Schulze:2016it, Shibl:2017jl}
The structure of the ML-MCTDH wavefunction is related to the tree
tensor network state,\cite{Shi:2006hz} which is a generalization of
the MPS. In fact, a different form of the equation of motion for the
MPS-MCTDH wavefunction ansatz can be derived in the framework of the
multi-layer formulation
(see Appendix A and B).
%
%}

\section{Theory}
In this work, a MPS formulation of the MCTDH theory (MPS-MCTDH) is
presented. It is expected to be applied to extended molecular systems
where many phonons are strongly and complexly correlated {\it via} the
electronic inter-state couplings as is often the case with molecular
systems.
In the MPS-MCTDH method, the molecular wavefunction is parametrized as
\begin{align}
\ket{\Psi} = \sum_\al \ket{\al} \ket{\Psi^\al} = \sum_\al \ket{\al} \sum_J A^\al_J \ket{\Phi^\al_J},\label{eq:wf_mps_1}
\end{align}
where
\begin{align}
%  A^{(\al)}_{J}
  A^{(\al)}_{J(=\{j_1,\cdots,j_f\})}
%  A^{(\al)}_{j_1 \cdots j_f}
              &\equiv \sum_{\ta_1\cdots\ta_{f-1}}
              a^{j_1}_{\ta_1} a^{j_2}_{\ta_1 \ta_2} %a^{j_3}_{\ta_2 \ta_3}
              \cdots %a^{j_{f-1}}_{\ta_{f-2} \ta_{f-1}}
              a^{j_f}_{\ta_{f-1}},
%  A^{(\al)}_J &\equiv \sum_{\ta_1\ta_2\cdots\ta_{k-1}}
%              a^{n_1}_{\ta_1} a^{n_2}_{\ta_1 \ta_2}
%              \cdots a^{n_{k-1}}_{\ta_{k-2} \ta_{k-1}} a^{n_k}_{\ta_{k-1}},\\
%  n_p &= (j_{n_xx},\cdots,j_{xx'})
  \label{eq:wf_mps_2}
\\
%\end{align}
%\begin{align}
%  \text{and}\;\;\;\;
  \ket{\Phi^{\al}_{J}}
%  \ket{\Phi^{\al}_{J(=\{j_1,\cdots,j_f\})}}
  &\equiv \ket{\vphi^\text{(1)}_{j_1} \vphi^{(2)}_{j_2} \cdots \vphi^{(f)}_{j_f}},
  \label{eq:wf_mps_3}
\end{align}
$\al$ denotes an electronic states in the multiset formalism and thus
satisfies $\braket{\al|\be} = \de_{\al\be}$ \cite{Worth:1996ck},
$a^{j_p}_{\ta_{p-1} \ta_p}$ is a site function of the MPS, and
$\vphi^{(p)}_{j_p}$ is a site basis states which is expressed as a
linear combination of the primitive functions as
$\sum_{r}c_{jr}\chi^p_r(Q_p)$ where $c_{jr}$ are variational
parameters as well as the site functions $a^{j_p}_{\ta_{p-1} \ta_p}$.
In constast to the linear coefficients $A^{(\al)}_{J}$, of which the
dimension grows rapidly as $O(n^f)$ where $n$ is the number
of the site basis per site $j_p = 1\;\cdots\;n$, the dimension of the
MPS site functions $\{a^{j_p}_{\ta_{p-1}\ta_{p}}\}$ grows as only
$O(nm^2f)$ where $m$ is the bond dimension of each site
functions, $\ta_{p} = 1\;\cdots\;m$. Note that the high-dimensional
$A^{(\al)}_{J}$ ($O(n^f)$) are never explicitly constructed
in the MPS-MCTDH method.
In the current implementation, it is possible to combine several
phonon modes into single site in the MPS, similarly to the multimode
single-particle function $Q_p=(q_i,q_j,\cdots)$ of the MCTDH method.
%

%about the tangent-space projector
Time-evolution of the variational parameters based on the TDVP is
formulated by using the tangent space projector $\hat{\cal
  P}_{[\Psi]}$
\begin{align}
  \ket{\dot{\Psi}} = -i \hat{\cal P}_{[\Psi]} \hat{H} \ket{\Psi}.
\end{align}
The tangent space projector acts as orthogonal projection for an
arbitrary vector onto the tangent plane, i.e. within the variational
space, at the current point $\Psi(t)$; thus, the time-dependent
Schr\"{o}dinger equation is satisfied at the first-order with respect
to all the variational parameters. The tangent space projector is
expressed as a summation of subspace projectors that must be
orthogonal to each other.
As the subspace projectors for the MPS site coefficients, the author
adopted the projector that imposes the left-gauge fixing condition
developed in Ref[\onlinecite{Haegeman:2016fo}]
%\begin{align}
%  \hat{\cal P}_\text{MPS} &=
%%  \ket{\Psi^\Ltxx_{\ta_{p-1}}} \ket{\vphi^{(p)}_{j_p}} \ket{\Psi^\Rtxx_{\ta_{p  }}}
%  %  \bra{\Psi^\Rtxx_{\ta_{p  }}} \bra{\vphi^{(p)}_{j_p}} \bra{\Psi^\Ltxx_{\ta_{p-1}}}
%  \sum^{f}_{p=1}
%  \sum_{\ta_{p-1} j_p \ta'_{p}}
%  \ket{\Psi^\Ltxx_{\ta _{p-1}}\;\vphi^{(p)}_{j_p}\;\Psi^\Rtxx_{\ta'_{p  }}}
%  \bra{\Psi^\Rtxx_{\ta'_{p  }}\;\vphi^{(p)}_{j_p}\;\Psi^\Ltxx_{\ta _{p-1}}}\nonumber\\
%  & -
%  \sum^{f-1}_{p=1}
%  \sum_{\ta_{p} \ta'_{p}}
%  \ket{\Psi^\Ltxx_{\ta _{p  }}\;\Psi^\Rtxx_{\ta'_{p  }}}
%  \bra{\Psi^\Rtxx_{\ta'_{p  }}\;\Psi^\Ltxx_{\ta _{p  }}},
%\end{align}
\begin{align}
  \hat{\cal P}_\text{\tiny MPS}
  &= \sum^{f     }_{p=1} \hat{\cal P}^+_{p}
   - \sum^{f\dash}_{p=1} \hat{\cal P}^-_{p},\label{eq:tang_proj}\\
%\end{align}
%where
%\begin{align}
  \hat{\cal P}^+_{p}
  &\equiv \sum_{\la_{p-1} j_p \la_{p}}
  \ket{\Psi^\Ltxx_{\la_{p-1}}\;\vphi^{(p)}_{j_p}\;\Psi^\Rtxx_{\la_{p  }}}
  \bra{\Psi^\Rtxx_{\la_{p  }}\;\vphi^{(p)}_{j_p}\;\Psi^\Ltxx_{\la_{p-1}}},\nonumber\\
  \hat{\cal P}^-_{p}
  &\equiv \sum_{\ga_{p} \la_{p}}
  \ket{\Psi^\Ltxx_{\ga_{p  }}\;\Psi^\Rtxx_{\la_{p  }}}
  \bra{\Psi^\Rtxx_{\la_{p  }}\;\Psi^\Ltxx_{\ga_{p  }}},\nonumber
\end{align}
and
\begin{align}
  \ket{\Psi^\Ltxx_{\ta_{p  }}}&\equiv \sum_{\ta_{1}\cdots\ta_{p-1}}
  L^{j_1}_{\ta_1} \cdots L^{j_{p  }}_{\ta_{p-1} \ta_{p  }}
  \ket{\vphi^{(1)}_{j_1} \cdots \vphi^{(p  )}_{j_{p  }}},
%  \ket{\Psi^\Ltxx_{\ta_{p-1}}}&\equiv
%  L^{j_1}_{\ta_1} \cdots L^{j_{p-1}}_{\ta_{p-2} \ta_{p-1}}
%  \ket{\vphi^{(1)}_{j_1} \vphi^{(2)}_{j_2} \cdots \vphi^{(p-1)}_{j_{p-1}}}
  \label{eq:lbasis}\\
  \ket{\Psi^\Rtxx_{\ta'_{p  }}}&\equiv\sum_{\ta_{p+1}\cdots\ta_{f-1}}
  R^{j_{p+1}}_{\ta'_{p} \ta_{p+1}} \cdots R^{j_f}_{\ta_{f-1}}
  \ket{\vphi^{(p+1)}_{j_{p+1}} \cdots \vphi^{(f)}_{j_{f}}},\label{eq:rbasis}
\end{align}
where $L^{j_{p-1}}_{\ta_{p-2}\ta_{p-1}}$ and $R^{j_{p+1}}_{\ta_p
  \ta_{p+1}}$ denote the left-orthonormal and right-orthonormal site
functions, respectively, appearing in the $p$-canonical form of MPS
wavefunctions
\begin{align}
  A^{(\al)}_{J} =
  \sum_{\ta_1\cdots\ta_{f-1}}
  L^{j_1}_{\ta_1}
  \cdots
  L^{j_{p-1}}_{\ta_{p-2} \ta_{p-1}}
  C^{j_{p  }}_{\ta_{p-1} \ta_{p  }}
  R^{j_{p+1}}_{\ta_{p  } \ta_{p+1}}
  \cdots
  R^{j_f}_{\ta_{f-1}},\nonumber
\end{align}
which can be transformed to the next site ($p$+$1$)-canonical form by
using the relation
\begin{align}
  \sum_{\ta'_{p }}
  C^{j_{p  }}_{\ta_{p-1} \ta'_{p }}
  R^{j_{p+1}}_{\ta'_{p } \ta_{p+1}}
&= \sum_{\ta'_{p } \ta_{p  }}
  L^{j_{p  }}_{\ta_{p-1} \ta_{p  }}
  \si_{\ta_{p  } \ta'_{p }}
  R^{j_{p+1}}_{\ta'_{p  } \ta_{p+1}} \nonumber\\
&= \sum_{\ta_{p }}
  L^{j_{p  }}_{\ta_{p-1} \ta_{p  }}
  C^{j_{p+1}}_{\ta_{p  } \ta_{p+1}}, \label{eq:gauge_trf}
\end{align}
where
\begin{align}
%\Biggl(
  \si_{\ta_{p  } \ta'_{p }}
\equiv
  \sum_{j'_{p  } \ta'_{p-1}}
  L^{j'_{p  }}_{\ta'_{p-1} \ta_{p  }}
  C^{j'_{p  }}_{\ta'_{p-1} \ta'_{p }},
%\Biggr)\nonumber
\end{align}
because $L^{j _{p  }}_{\ta _{p-1} \ta_p}$ is obtained by the diagonalization of
\begin{align}
%   ^\text{\tiny (L)}\Gamma^{\ta'_{p-1} j'_{p  }}_{\ta_{p-1} j_{p  }}
%&\equiv
  \sum_{\ta_p} C ^{j'_{p  }}_{\ta'_{p-1} \ta_p}
          \ovl{C}^{j _{p  }}_{\ta _{p-1} \ta_p}
= \sum_{\ta_p} L ^{j'_{p  }}_{\ta'_{p-1} \ta_p} w^\text{\tiny (L)}_{\ta_p}
          \ovl{L}^{j _{p  }}_{\ta _{p-1} \ta_p},\label{eq:LwL}
%%\\
%%%   ^\text{\tiny (R)}\Gamma^{j'_{p+1} \ta'_{p+1}}_{j_{p+1} \ta_{p+1}}
%%%&\equiv
%%  \sum_{\ta_p} C ^{j'_{p+1}}_{\ta_p \ta'_{p+1}}
%%          \ovl{C}^{j _{p+1}}_{\ta_p \ta _{p+1}}
%%= \sum_{\ta_p} R ^{j'_{p+1}}_{\ta_p \ta'_{p+1}} w^\text{\tiny (R)}_{\ta_p}
%%          \ovl{R}^{j _{p+1}}_{\ta_p \ta _{p+1}}
\end{align}
for details refer to literature, e.g. Ref[\onlinecite{Schollwoeck:2010gl}].
In the case of the MPS-MCTDH ansatz, this projector generates not only
the variation of site functions of the MPS, but also the variation of
site basis states, themselves. It should bring a complication for its
formulation, thus a certain MCTDH gauge,
$\braket{\vphi^{(p)}_j|\dot{\vphi}^{(p)}_l}=0$,
%$\braket{\vphi^{(p)}_j|\vphi^{(p)}_l}=\de_{jl}$
is adopted, by which the variation of the site basis states generated
by $\hat{\cal P}_\text{MPS}$ are vanished. The projector for the
complementary space $\hat{{\cal P}'} \equiv \hat{\cal P} - \hat{\cal
  P}_\text{MPS}$, i.e.\ the site basis functions space, can be derived
in the same way as Ref[\onlinecite{Lubich:2015fr, Kloss:2017kq,
    Bonfanti:2018vj}] except that the MCTDH coefficient is replaced by
the MPS, and the
%
%\textcolor{red}{symmetric}
%
Lie-Trotter splitting of
the tangent space projectors should be done naturally for all the
projector
\begin{align}
  e^{-i\hat{\cal P}\hat{H}2\de} \approx
  e^{-i\hat{{\cal P}'}\hat{H}\de}
  e^{-i\hat{\cal P}_\text{MPS}\hat{H}\de}
  e^{-i\hat{{\cal P}'}\hat{H}\de}.
\end{align}

Another way is to assume the constant mean-field (CMF)
approximation,\cite{Beck:1997fs} in which the mean-field operators
and the integrals written in the site basis functions are frozen
during each step of the propagation, and thus the time-evolution of
the MPS site functions and the site basis functions are decoupled
during the step intervals.
The CMF integration method, which was adopted in this work, becomes
advantageous for extended systems where the computation of the
mead-field operators and related operations are the most
time-consuming steps.

Another issue of discussion for using the MPS form in the MCTDH theory
is the construction of mean-field operators for single-particle
functions $\{\vphi^{(p)}_{j_p}\}$ expressed as
\begin{align}
  \braket{\hat{O}}^{\al\be(p)}_{jk} = \bra{\Psi^{\al(p)}_j} \hat{O} \ket{\Psi^{\be(p)}_k},
%  \sum_{j_1} \cdots \sum_{j_{p-1}}\sum_{j_{p+1}} \cdots \sum_{j_f}
%  \ovl{A}_{j_1 \cdots j_{p-1} j j_{p+1}}                           \nonumber\\
%  \sum_{k_1} \cdots \sum_{k_{p-1}}\sum_{k_{p+1}} \cdots \sum_{k_f}
%       A _{k_1 \cdots k_{p-1} k k_{p+1}}                           \nonumber\\
%  \bra{\vphi_{j_1} \cdots \vphi_{j_{p-1}} \vphi_{j_{p+1}} \cdots \vphi_{j_f}}
%  \hat{O}(Q_1, \cdots , Q_f)
%  \ket{\vphi_{k_1} \cdots \vphi_{k_{p-1}} \vphi_{k_{p+1}} \cdots \vphi_{k_f}}
\end{align}
where
\begin{align}
  \ket{\Psi^{\be(p)}_k} \equiv
  \sum_{k_1} \cdots \sum_{k_{p-1}}\sum_{k_{p+1}} \cdots \sum_{k_f}
       A^{(\be)} _{k_1 \cdots k_{p-1} k k_{p+1}}                           \nonumber\\
  \times
  \ket{\vphi^{(1)}_{k_1} \cdots \vphi^{(p-1)}_{k_{p-1}} \;
       \vphi^{(p+1)}_{k_{p+1}} \cdots \vphi^{(f)}_{k_f}}.
\end{align}
In general, phonon modes of molecular systems are complexly coupled
through the potential energy surface created by the electronic-state;
thus, there can be a $f$-body interaction $\hat{V}(Q_1,\cdots,Q_f)$
term in the Hamiltonian, but in many cases the $f$-body interaction is
efficiently expanded by the $n$-mode coupling
representation \cite{Carter:1997ej} and usually it is sufficient to
truncate up to the fourth order expansion \cite{Yagi:2004jy}. An
efficient evaluation of three and four-body operators for MPS
wavefunction by the DMRG sweep algorithm is presented in our previous
work \cite{Kurashige:2011ck}. Alternatively, the $f$-body Hamiltonian
are reduced to products of one-body operators of the single site basis
$\vphi^{(p)}(Q_p)$ in the MCTDH method \cite{Beck:2000wm} as
\begin{align}
  &\bra{\vphi^{(1)}_{j_1} \cdots \vphi^{(f)}_{j_f}} \hat{O}^\text{(1,..,$f$)}
   \ket{\vphi^{(1)}_{k_1} \cdots \vphi^{(f)}_{k_f}} \nonumber\\
  &= \sum_{a} c_a \bra{\vphi^{(1)}_{j_1}} \hat{o}_a^\text{(1)}  \ket{\vphi^{(1)}_{k_1}}
           \cdots \bra{\vphi^{(f)}_{j_f}} \hat{o}_a^\text{($f$)}\ket{\vphi^{(f)}_{k_f}}.
\end{align}
This product form is also very suitable for the MPS wavefunction.
%because
%
For example, a mean-field operator in the product form is decomposed
as
\begin{align}
  \braket{\hat{O}_a}^{(p)}_{jk}
  &= \ovl{C}^{j}_{\la_{p-1} \la_{p}}
          \bra{\Psi^\Ltxx_{\la_{p-1}}} \hat{O}^\text{(1,..,$p-$1)}_a   \ket{\Psi^\Ltxx_{\ta_{p-1}}} \nonumber\\
  &\times \bra{\Psi^\Rtxx_{\la_{p  }}} \hat{O}^\text{($p$+1,..,$f$)}_a \ket{\Psi^\Rtxx_{\ta_{p  }}}
          C ^{k}_{\ta_{p-1} \ta_{p}}
          \hat{o}^{(p)}_a.\label{eq:mfop}
\end{align}
The operators in the left and right blocks can be prepared easily in
the same way as the DMRG algorithm, and the mean-field operators for
the different sites $p$ are constructed by the $p$-canonical form
transformation along the MPS lattice.

\section{Results and discussion}
To demonstrate the efficiency of the MPS-MCTDH method, exciton-phonon
dynamics in the molecular aggregates were performed.
%other related MCTDH applications for excitonic systems
The applications to the exciton-phonon dynamics are important for
computing the quantum efficiency of energy conversion and transfer in
molecular aggregates,\cite{Tamura:2007ke, Tamura:2012fv,
  Tamura:2015il, Binder:2017hv} e.g.\ photovoltaic and photosynthetic
systems, but application to the extended systems were hampered by the
exponential increase of the computational cost with the number of
vibrational degree of freedoms.
Most recently, Ren and co-worker successfully computed the absorption
and fluorescence spectra of molecular aggregates with a time-dependent
DMRG algorithm at both zero and finite temperature.\cite{Ren:2018tf}
%Holstein Hamiltonian
The Hamiltonian adopted for the benchmark simulation is
\begin{align} %\mathbbm{1}
  \hat{H} &= \sum_{\al} \ket{\al}\bra{\al} \sum_{v} \frac{\omega^\al_v}{2}
  \left(- \frac{\pa^2}{\pa Q_v^2} + \hat{Q}_v^2 \right) \nonumber\\
  &+ \sum_{\al} \ket{\al}\bra{\al} \sum_{v} \hat{Q}_v \ka^\al_v
  + \sum_{\al\be} \ket{\al}\bra{\be} J_{\al\be},\label{eq:ham_exiton}
\end{align}
where $\omega^\al_v$ is the harmonic frequency of the phonon mode $v$
on the diabatic electronic-state $\al$, $\ka^\al_v$ is the first-order
coupling between the diabatic electronic-state $\al$ and the phonon
mode $v$, and $J_{\al\be}$ for $\al\neq\be$ is the diabatic coupling
between $\al$ and $\be$ states and that for $\al$=$\be$ is the
energy gaps between the states at the origin, $Q_v$=$0$ for all $v$.
%${\mathbbm Q}=0$.
%
The diabatic electronic-state of the molecular aggregate, $\ket{\al}$,
is characterized by the electronic state of each molecules; if the
$i$-th molecule is in its excited-state, e.g.\ $S_1$ state, while all
the other molecules are in the ground state, the state is denoted by
$\ket{i}$.
% parameters
The parameters in the Hamiltonian, $\omega^\al_v$, $\ka^\al_v$, and
$J_{\al\be}$, can be quantitatively determined by {\it ab initio}
quantum chemical calculations to simulate real molecular systems.
%In this study, XXXXX parameters were used.
%
In the following, a single local vibrational mode per monomer is
considered; thus the number of phonon DOFs is equal to the number of
molecules and to the number of electronic states. The dimension for
the site basis functions is set to four for all the phonon DOFs, and
each site basis function is expressed by a linear combination of eight
eigenfunctions of the harmonic oscillator from the lowest quantum
number.
The MCTDH and MPS-MCTDH methods described in this work were
implemented in Python3.

\begin{figure}[b]
%  \centering
  \includegraphics[width=8.8cm]{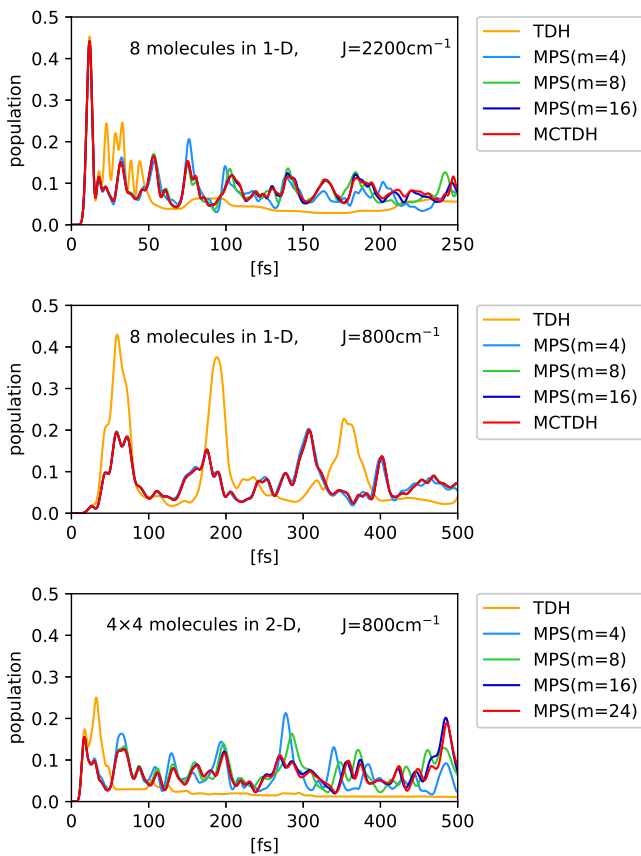}
  \caption{\label{fig:MOL-8_J-2200} Time evolution of the exciton
    population with various methods for the 1-D and 2-D systems. The
    populations at the opposite end of the initially populated site
    are shown.}
\end{figure}

Figure~\ref{fig:MOL-8_J-2200} shows the population dynamics of an
exciton in a molecular aggregate consisting of eight molecules 1-D
aligned and sixteen molecules 2-D aligned.
%in weak and strong electronic-couping regime
The population of the electronic-state $\ket{i}$ in which the exciton
is localized on the $i$-th molecule from the end is calculated as
$\rho_{i}(t) = \braket{\Psi|i}\braket{i|\Psi} =
\braket{\Psi^i|\Psi^i}$. At time $t=0$, only the monomer at one end of
the 1-D aggregate is electronically excited, and time evolution of the
population of the exciton on the opposite end monomer are shown.
%parameters
The values for the parameters are $\omega^\al_v$=1255 cm$^{-1}$ and
$\ka^\al_v$=1.072$\times \omega^\al_v$, typical values for
intramolecular mode of organic semiconductors. The electronic states
in which the excitons are located at the nearest neighbor to each
other interact with each other by the diabatic coupling $J$ in 1-D and
2-D systems.
%

%\begin{figure}%[htbp]

%(0)TDH
A multiset time-dependent Hartree (TDH) method, in which each vibrational
wavefunction $\Psi^\al$ associated with an electronic-state is expressed
by a Hartree product, hence the vibrational DOFs are non-entangled,
are also performed for comparison. In all cases, the deviations
between the TDH and MCTDH methods are found to be significant, which
can be regarded as a measure of the strength of the entanglement
between the vibrational DOFs in the system.
%(1)the first wave arrival dependence on the strength of J
The earliest maximum of the population in the plots corresponds to the
first arrival of the exciton from the initial location, i.e.\ reflects
the mobility of excitons, and the waves should go back and forth
causing interference with each other.
For the 1-D system with the $J$=800 cm$^{-1}$ (top panel), the MPS well
reproduces the results of the MCTDH method, even with the smallest
bond dimensions, $m$=4.
For the same 1-D system with $J$=2200 cm$^{-1}$ (middle
panel), the MPS($m$=4) plots deviates from the MCTDH plots after 70 fs
for. The MPS($m$=8) is more robust and can trace the MCTDH plot and
slightly deviate after 200 fs. The MPS with the largest bond dimensions
$m$=16 reproduces the MCTDH result. It suggests that the required size
of the bond dimension $m$ to maintain the same accuracy depends on the
strength of the coupling.
For the 2-D system with $J$=800 cm$^{-1}$ (bottom panel), the
convergence of population dynamics to the result with large $m$ is
slower than that for the 1-D system with the same coupling strength.
% $m$=16 is required to converge.
Due to the nature of the MPS, which is a sequential product of
tensors, the required size of $m$ is increased for 2-D and higher
dimensional systems.

\begin{table}
\caption{\label{tab:timing} Elapsed CPU time (sec) for computing the
  MPS related operations, i.e.\ computing the mean-field operators and
  propagation of the MPS site functions, in a step of the CMF
  integration algorithm. The timings for the corresponding operations
  in the MCTDH method are also shown.}
\begin{ruledtabular}
\begin{tabular}{cccccccccccccccc}
\# of    &MCTDH& \multicolumn{4}{c}{MPS-MCTDH method}\\\cline{3-6}
molecules&
        method &$m=4$&$m=8$&$m=16$&$m=32$\\
\hline\\[-10pt]
% 4   &  0.1  & 0.3 & 0.3 & $-$  & $-$ \\
  8   &  5.9  & 1.1 & 1.4 &  2.4 &  5.9\\
 10   &  224  & 1.8 & 2.2 &  4.2 &   12\\
 12   &16907  & 2.5 & 3.3 &  6.8 &   22\\
 16   & $-$   & 4.8 & 6.7 &   14 &   53\\
 24   & $-$   &  13 &  19 &   45 &  179\\
 32   & $-$   &  27 &  42 &   99 &  421\\
 64   & $-$   & 270 & 415 & 1013 & 4638\\
 96   & $-$   & 859 &1373 & 3372 &15509
\end{tabular}
\end{ruledtabular}
\end{table}

Table~\ref{tab:timing} shows the elapsed CPU time for computing the
mean-field operators and propagation of the MCTDH coefficients $A_{J}$
or the MPS site functions $a^{j_1}_{\ta_1}, a^{j_2}_{\ta_1 \ta_2},
\cdots, a^{j_f}_{\ta_{f-1}}$, by the short-time Lanczos algorithm, per
one step of the CMF integration algorithm, in which the mean-filed
operators are evaluated two times and the coefficient are propagated
three times including the backward propagation for the error
estimation \cite{Beck:1997fs}.
Due to the linear parametrization of the Hilbert space, the elapsed
time for the MCTDH grows exponentially with the number of the
molecules.
The formal computational scaling of the MPS is $O(n m^3 k^2
f)$, where $k$ is the number of electronic states, for the evaluation
of the mean-field operators and the time propagation of the site
functions. In this excitonic systems, the number of vibrational DOFs
$f$ and electronic states $k$ grows linearly with the number of
molecules $N$, thus the formal computational scaling with respect to
$N$ is cubic when $n$ and $m$ are constant, and that with $m$ is also
cubic. The actual scaling with $m$ between $m$=8 and $m$=16 is
linear, and that between $m$=16 and $m$=32 is quadratic, i.e.\ smaller
than the formal scaling. This is because the most time consuming steps
are relatively small matrix-matrix multiplication; typically the tensor
contraction in Eq.(\ref{eq:mfop}), in which the matrix size for the
matrix-matrix multiplication is $m\times m$, and high throughput
cannot be achieved by the current implementation for small $m$. %XXXXX
%Conclusions.
%\section{Summary}%

To summarize, A matrix product state formulation of the
multiconfiguration time-dependent Hartree is presented.
The MPS can efficiently encode the Hilbert space of the phonon DOFs,
which grows exponentially with the number of modes. Due to the nature
of the MPS form, it is particularly effective for extended systems and
modes that are strongly correlated as is often the case with molecular
systems.
The efficiency of the method was demonstrated on the quantum dynamics
of the extended excitonic systems.

\section*{Acknowldgements}
The author thanks Yoshitaka Tanimura, Tatsushi Ikeda, and Takeshi Sato
for many valuable discussions.

%Appendix
\section*{Appendix A: Algorithm for propagating the MPS site coefficients}
%In the calculations of this study, the MPS site coefficients are
%propagated in time as follow.
%when imporse the left-gauge fixing condition
The MPS site coefficients are propagated sequentially from the
leftmost site $p$=1 to the rightmost site $p$=$f$, in the former half,
and then proceeds in the opposite direction, i.e.\ from the site
$p$=$f$ to the site $p$=1 in the latter half of the
algorithm, to accomplish the time propagation written by
\begin{align}
  &\ket{\Psi\text{\small (t+2$\de$)}}
  =
  e^{-i\hat{\cal P}_\text{\tiny MPS}\hat{H}2\de}\ket{\Psi\text{\small (t)}}\nonumber\\
  &=
  e^{-i(\hat{\cal P}^+_1 - \hat{\cal P}^-_1
%       +\hat{\cal P}^+_2 - \hat{\cal P}^-_2
       +\cdots
%       +\hat{\cal P}^+_{f\dash} - \hat{\cal P}^-_{f\dash}
       - \hat{\cal P}^-_{f\dash}
       +2\hat{\cal P}^+_f
       -\hat{\cal P}^-_{f\dash}
       + \cdots
       -\hat{\cal P}^-_{1} + \hat{\cal P}^+_{1}
       )\hat{H}\de}\ket{\Psi\text{\small (t)}}\nonumber
  \\
  &\approx
  e^{-i\hat{\cal P}^+_1\hat{H}\de}
  e^{+i\hat{\cal P}^-_1\hat{H}\de}
  \cdots
  e^{+i\hat{\cal P}^-_{f\dash}\hat{H}\de}
  e^{-i\hat{\cal P}^+_{f     }\hat{H}\de}\nonumber\\
  &\;\times
  e^{-i\hat{\cal P}^+_{f     }\hat{H}\de}
  e^{+i\hat{\cal P}^-_{f\dash}\hat{H}\de}
  \cdots
  e^{+i\hat{\cal P}^-_1\hat{H}\de}
  e^{-i\hat{\cal P}^+_1\hat{H}\de}
  \ket{\Psi\text{\small (t)}}.
\end{align}
The time integrations based on the symmetric expansion are expected to
have a favorable error of order ${\cal O}(\de^3)$. In the following,
the index $\al$ that denotes electronic states is omitted for clarity.
{\setlength{\leftmargini}{12pt}
\begin{enumerate}[listparindent=\parindent]%[label=\textbf{\arabic*}.]
  \renewcommand{\labelenumi}{\arabic{enumi})}
  \setcounter{enumi}{-1}
\item At the beginning, the MPS wavefunction is prepared in the
  [$p$=1]-canonical form, i.e.\ the coefficients except for the
  site $p$=1 are right-orthonormal,
  \begin{align}
%    \ket{\Psi} =
    \sum_{{\bm \ta},J}
    C^{j_1}_{\ta_1}
    R^{j_2}_{\ta_1 \ta_{2}}
    \cdots
    R^{j_{f\dash}}_{\ta_{f-2} \ta_{f\dash}}
    R^{j_f}_{\ta_{f\dash}}
    \ket{\Phi_J},\nonumber
  \end{align}
  which is always available through the gauge transformation by
  Eq.(\ref{eq:gauge_trf}) if necessary.
\item In this step, the coefficients for the current site (starts with
  $p$=1), $C^{j_p}_{\ta_{p\dash} \ta_{p}}$, are propagated by applying
  the operator exponential $e^{-i\hat{\cal P}^+_p\hat{H}\de}$
%->  where
%->  \begin{align}
%->    \hat{\cal P}^\text{+}_p =
%->    \sum_{\la_{p\dash} l_p \la_{p}}
%->    \ket{\Psi^\Ltxx_{\la_{p\dash}}\;\vphi^{(p)}_{l_p}\;\Psi^\Rtxx_{\la_{p }}}
%->    \bra{\Psi^\Rtxx_{\la_{p }}\;\vphi^{(p)}_{l_p}\;\Psi^\Ltxx_{\la_{p\dash}}},\nonumber
%->  \end{align}
%->  which is one of the terms in the first summation of
%->  Eq.(\ref{eq:tang_proj}),
  to the wavefunction as
  \begin{align}
    C'^{j_p}_{\ta_{p\dash} \ta_{p}}
%    &\equiv
%    C^{j_p}_{\ta_{p\dash} \ta_{p}}\text{\small $(t+\de/2)$}\nonumber\\
    &=
    \bra{\Psi^\Rtxx_{\ta_{p }}\;\vphi^{(p)}_{j_p}\;\Psi^\Ltxx_{\ta_{p\dash}}}
    e^{-i\hat{\cal P}^+_p\hat{H}\de} \ket{\Psi}\nonumber\\
    &=
    \bra{\Psi^\Rtxx_{\ta_{p }}\;\vphi^{(p)}_{j_p}\;\Psi^\Ltxx_{\ta_{p\dash}}}
    e^{-i\hat{\cal P}^+_p\hat{H}\de} \nonumber\\
    &\phantom{=}\times \sum_{\ta'_{p\dash} j'_p \ta'_{p}}
    C^{j'_p}_{\ta'_{p\dash} \ta'_{p}}%\text{\small $(t)$}
    \ket{\Psi^\Ltxx_{\ta'_{p\dash}}\;\vphi^{(p)}_{j'_p}\;\Psi^\Rtxx_{\ta'_{p }}}.
  \end{align}
  It is expressed as
  \begin{align}
%  {\bm c}\text{\small $(t+\de/2)$}
%  = e^{-i {\bm H}\de/2}\;{\bm c}\text{\small $(t)$}\label{eq:expH}
  {\bm c}'
  = e^{-i {\bm H}\de}\;{\bm c}\label{eq:expH}
  \end{align}
  where the indices $(\ta_{p\dash},j_p,\ta_{p})$ of
  $C^{j_p}_{\ta_{p\dash} \ta_{p}}$ are flattened to one dimension in
  the vector ${\bm c}$ and ${\bm H}$ is a matrix defined as
  \begin{align}
    H^{(\ta_{p\dash},j_p,\ta_{p})}_{(\ta'_{p\dash}j'_p,\ta'_{p})}
    \equiv
    \bra{\Psi^\Rtxx_{\ta_{p }}\;\vphi^{(p)}_{j_p}\;\Psi^\Ltxx_{\ta_{p\dash}}}
    \hat{H}
    \ket{\Psi^\Ltxx_{\ta'_{p\dash}}\;\vphi^{(p)}_{j'_p}\;\Psi^\Rtxx_{\ta'_{p }}}.
  \end{align}

  The multiplication of the matrix exponential in Eq.(\ref{eq:expH})
  is evaluated by an efficient short iterative Lanczos (SIL)
  integrator, which is one of the Krylov subspace methods and requires
  only the matrix-vector multiplications ${\bm v} = {\bm H}{\bm c}$.
  Analogous to the quantum chemical DMRG, of which the Hamiltonian
  contains four-site operators,\cite{Kurashige:2009gs} the Hamiltonian
  is decomposed as
  \begin{align} \hat{H} = \sum_{i} \hat{\cal O}^i_\Ltxx \hat{\cal O}^i_\Ctx \hat{\cal O}^i_\Rtxx\label{eq:op_lcr}
%           (\equiv \sum_{i} \hat{\cal O}^i_{\LCRtxx})
  \end{align}
  where the operator $\hat{\cal O}^i_\Ltxx$, $\hat{\cal O}^i_\Ctx$, and
  $\hat{\cal O}^i_\Rtxx$ act on the phonon modes $Q_1 \cdots Q_{p\dash}$,
  $Q_{p}$, and $Q_{p+1} \cdots Q_{f}$, respectively.
  Now the Hamiltonian multiplication is reduced to
  \begin{align}
    v^{j_p}_{\ta_{p\dash} \ta_p}
    &= \sum_{\ta'_{p\dash},j'_p,\ta'_p} H^{\ta_{p\dash},j_p,\ta_p}_{\ta'_{p\dash},j'_p,\ta'_p} C^{j'_p}_{\ta'_{p\dash},\ta'_p}\nonumber\\
    &=
    \sum_i \sum_{\ta'_{p\dash}} \braket{{\cal O}^i_\Ltxx}^{\ta_{p\dash}}_{\ta'_{p\dash}}
    \sum_{j'_p} \braket{{\cal O}^i_\Ctx}^{j_p}_{j'_p}
    \sum_{\ta'_p} \braket{{\cal O}^i_\Rtxx}^{\ta_p}_{\ta'_p} C^{j'_p}_{\ta'_{p\dash},\ta'_p},
  \end{align}
%
%  \begin{align}
%    v_{l,c,r}
%    &= \sum_{l',c',r'} H^{l,c,r}_{l',c',r'} C^{c'}_{l',r'}\nonumber\\
%    &=
%    \sum_i \sum_{l'} \braket{{\cal O}^i_\Ltxx}^{l}_{l'}
%    \sum_{c'} \braket{{\cal O}^i_\Ctx}^{c}_{c'}
%    \sum_{r'} \braket{{\cal O}^i_\Rtxx}^{r}_{r'} C^{c'}_{l',r'},
%  \end{align}
 %
  %here $l,r,c$ are used instead of $\ta_{p\dash}$, $j_p$, $\ta_{p}$ for simplicity,
  where
  \begin{align}
    \braket{{\cal O}^i_\Ltxx}^{\ta_{p\dash}}_{\ta'_{p\dash}}
    &\equiv
    \bra{\Psi^\Ltxx_{\ta_{p\dash}}}
    \hat{\cal O}^i_\Ltxx
    \ket{\Psi^\Ltxx_{\ta'_{p\dash}}},
    \nonumber\\
    \braket{{\cal O}^i_\Ctx}^{j_p}_{j'_p}
    &\equiv
    \bra{\vphi^{(p)}_{j_p}}
    \hat{\cal O}^i_\Ctx
    \ket{\vphi^{(p)}_{j'_p}},
    \nonumber\\
    \braket{{\cal O}^i_\Rtxx}^{\ta_{p}}_{\ta'_{p}}
    &\equiv
    \bra{\Psi^\Rtxx_{\ta_{p }}}
    \hat{\cal O}^i_\Rtxx
    \ket{\Psi^\Rtxx_{\ta'_{p }}}.
  \end{align}
  The computational scaling of this step is, therefore, $
  O(nm^3\til{k})$ where $n$,$m$, and $\til{k}$ are the dimension of
  $j_p$, $\ta_{p(\dash)}$, and $i$, respectively. Note that the
  summation over the phonon modes $v$ in the Hamiltonian
  [Eq.(\ref{eq:ham_exiton})] is not appeared in the summation over $i$
  in Eq.(\ref{eq:op_lcr}) because it has already been taken in the
  evaluation of $\braket{{\cal O}_{\Ltxx(\Rtxx)}}$, $e.g.$
  \begin{align}
    \sum^{f}_{v=1} g_v(\hat{Q}_v)
    &= \sum^{p\dash}_{v=1} g_v(\hat{Q_v}) \otimes \mathbbm{1}_\Ctx \otimes \mathbbm{1}_\Rtxx \nonumber\\
    &+ \mathbbm{1}_\Ltxx                   \otimes g_p(\hat{Q_p})   \otimes \mathbbm{1}_\Rtxx \nonumber\\
    &+ \mathbbm{1}_\Ltxx                   \otimes \mathbbm{1}_\Ctx \otimes \sum^{f}_{v=p+1} g_v(\hat{Q_v}),
  \end{align}
  thus $\til{k}$ grows only with the number of the electronic states
  $k$ ($c.f.$ Eq.(\ref{eq:ham_exiton})).
  \label{step:expH}
\item The wavefunction after the propagation
  $C^{j_p}_{\ta_{p\dash} \ta_{p}} \rightarrow C'^{j_p}_{\ta_{p\dash}
  \ta_{p}}$ in the previous step is expressed as
  \begin{align}
    %\ket{\Psi^\al} =
    %\sum_{\ta_1\cdots\ta_{f\dash},J}
    \sum_{{\bm \ta},J}
    L'^{j_1}_{\ta_1}
    \cdots
    L'^{j_{p\dash}}_{\ta_{p\text{-2}} \ta_{p\dash}}
   (C'^{j_{p  }}_{\ta_{p\dash} \ta_{p  }})
    R^{j_{p+1}}_{\ta_{p  } \ta_{p+1}}
    \cdots
    R^{j_f}_{\ta_{f\dash}} \ket{{\Phi^{\al}_{J}}}.\nonumber
  \end{align}
  It is transformed to
  \begin{align}
    %\ket{\Psi^\al} =
    %\sum_{\ta_1\cdots\ta_{f\dash},J}
    \sum_{{\bm \ta},\ga_p,J}
    L'^{j_1}_{\ta_1}
    \cdots
    L'^{j_{p\dash}}_{\ta_{p\text{-2}} \ta_{p\dash}}
   (L'^{j_{p  }}_{\ta_{p\dash} \ga_{p  }}
    \si'^{\ga_{p}}_{\ta_{p}})
    R^{j_{p+1}}_{\ta_{p  } \ta_{p+1}}
    \cdots
    R^{j_f}_{\ta_{f\dash}} \ket{{\Phi^{\al}_{J}}}.\nonumber
  \end{align}
  by using the orthogonal decomposition in Eq.(\ref{eq:gauge_trf})
%  \begin{align}
%    C'^{j_{p }}_{\ta_{p\dash} \ta_{p}}
%    =
%    \sum_{\ga_{p}}
%    L'^{j_{p  }}_{\ta_{p\dash} \ga_{p}}
%    \la_{\ga_{p}}
%    V_{\ga_{p} \ta_{p}}
%    =
%    \sum_{\ta'_{p}}
%    L'^{j_{p  }}_{\ta_{p\dash} \ga_{p}}
%    \si'^{\ga_{p}}_{\ta_{p}}\label{eq:QRforL}
%  \end{align}
  %
  \label{step:C2LS}
\item This step is similar to Step \ref{step:expH}, and the matrix
  $\si'_{\ga_{p} \ta_{p}}$ is propagated (but backward in time) by
  applying the operator exponential $e^{+i\hat{\cal P}^-_p\hat{H}\de}$
%  where
%  \begin{align}
%    \hat{\cal P}^{-}_p =
%    \sum_{\la_{p} \ga_{p}}
%    \ket{\Psi^\Ltxx_{\ga_{p  }}\;\Psi^\Rtxx_{\la_{p  }}}
%    \bra{\Psi^\Rtxx_{\la_{p  }}\;\Psi^\Ltxx_{\ga_{p  }}},\nonumber
%  \end{align}
%  which is one of the terms in the second summation of
%  Eq.(\ref{eq:tang_proj}),
  to the wavefunction as
  \begin{align}
    \si^{\ga_{p}}_{\ta_{p}}
%    &\equiv
%    C^{j_p}_{\ta_{p\dash} \ta_{p}}\text{\small $(t+\de/2)$}\nonumber\\
    &=
    \bra{\Psi^\Rtxx_{\ta_{p  }}\;\Psi^\Ltxx_{\ga_{p  }}}
    e^{+i\hat{\cal P}^{-}_p\hat{H}\de} \ket{\Psi}\nonumber\\
    &=
    \bra{\Psi^\Rtxx_{\ta_{p  }}\;\Psi^\Ltxx_{\ga_{p  }}}
    e^{+i\hat{\cal P}^{-}_p\hat{H}\de} \nonumber\\
    &\phantom{=}\times \sum_{\ga'_{p} \ta'_{p}}
    \si'^{\ga'_{p}}_{\ta'_{p}}%\text{\small $(t+\de/2)$}
    \ket{\Psi^\Ltxx_{\ga'_{p  }}\;\Psi^\Rtxx_{\ta'_{p  }}}.
  \end{align}
  It can be rewritten as
  \begin{align}
%  {\bm \si}\text{\small $(t)$}
%  = e^{+i {\bm K}\de/2}\;{\bm \si}\text{\small $(t+\de/2)$}\label{eq:expK}
  {\bm \si}
  = e^{+i {\bm K}\de}\;{\bm \si}'\label{eq:expK}
  \end{align}
  where the indices $(\ga_{p}, \ta_{p})$ of $\si'^{\ga_{p}}_{\ta_{p}}$
  are flattened to one dimension in the vector ${\bm \si}'$ and ${\bm
    K}$ is a matrix defined as
\begin{align}
    K^{(\ga_{p} \ta_{p})}_{(\ga'_{p} \ta'_{p})}
    \equiv
    \bra{\Psi^\Rtxx_{\ta_{p}}\;\Psi^\Ltxx_{\ga_{p}}}
    \hat{H}
    \ket{\Psi^\Ltxx_{\ga'_{p}}\;\Psi^\Rtxx_{\ta'_{p }}}.
  \end{align}
  The multiplication of the matrix exponential in Eq.(\ref{eq:expK})
  is evaluated by the SIL, which requires only the matrix-vector
  multiplication ${\bm u} = {\bm K}{\bm \si}$. As done in Step
  \ref{step:expH}, the Hamiltonian is decomposed as
  \begin{align}
    \hat{H} = \sum_{i} \hat{\cal O}^i_{\Ltxx} \hat{\cal O}^i_\Rtxx \label{eq:op_lr}
  \end{align}
  where the operator $\hat{\cal O}^i_{\Ltxx}$ and $\hat{\cal O}^i_\Rtxx$ act on the
  phonon modes $Q_1 \cdots Q_{p}$ and $Q_{p+1} \cdots Q_{f}$,
  respectively.
%  %As done in the {\it ab initio} DMRG in which four-index
  \begin{align}
    u^{\ga_p}_{\ta_p}
    &= \sum_{\ga'_p,\ta'_p} K^{\ga_p,\ta_p}_{\ga'_p,\ta'_p}
                            \si'^{\ga'_p}_{\ta'_p}\nonumber\\
    &=
    \sum_i \sum_{\ga'_p} \braket{{\cal O}^i_\Ltxx}^{\ga_p}_{\ga'_p}
    \sum_{\ta'_p} \lan {\cal O}^i_\Rtxx\ran^{\ta_p}_{\ta'_p} \si'^{\ga'_p}_{\ta'_p}.
  \end{align}
  %here $l,r,c$ are used instead of $\ta_{p\dash}$, $j_p$, $\ta_{p}$
  The computational scaling of this step is,
  therefore, $O(m^3\til{k})$.
  \label{step:expK}
\item The wavefunction obtained by the propagation
  $\si'^{\ga_{p}}_{\ta_{p}} \rightarrow \si^{\ga_{p}}_{\ta_{p}}$ in
  the previous step is expressed as
  \begin{align}
    %\ket{\Psi^\al} =
    %\sum_{\ta_1\cdots\ta_{f\dash},J}
    \sum_{{\bm \ta},\ga_p,J}
    L'^{j_1}_{\ta_1}
    \cdots
    L'^{j_{p}}_{\ta_{p\dash} \ta_{p}}
   (\si^{\ta_{p}}_{\ga_{p}}
    R^{j_{p+1}}_{\ga_{p  } \ta_{p+1}})
    R^{j_{p+2}}_{\ta_{p+1} \ta_{p+2}}
    \cdots
    R^{j_f}_{\ta_{f\dash}} \ket{{\Phi^{\al}_{J}}}.\nonumber
  \end{align}
  It is transformed to
  \begin{align}
    %\ket{\Psi^\al} =
    %\sum_{\ta_1\cdots\ta_{f\dash},J}
    \sum_{{\bm \ta},J}
    L'^{j_1}_{\ta_1}
    \cdots
    L'^{j_{p}}_{\ta_{p\dash} \ta_{p}}
   (C^{j_{p+1}}_{\ta_{p  } \ta_{p+1}})
    R^{j_{p+2}}_{\ta_{p+1} \ta_{p+2}}
    \cdots
    R^{j_f}_{\ta_{f\dash}} \ket{{\Phi^{\al}_{J}}}.\nonumber
  \end{align}
  \label{step:SR2C}
\item Steps \ref{step:expH}--\ref{step:SR2C} are repeated until the
  current site $p$ reaches the rightmost site $p$=$f$
  \begin{align}
%    \ket{\Psi} =
    \sum_{{\bm \ta},J}
    L'^{j_1}_{\ta_1}
    L'^{j_2}_{\ta_1 \ta_{2}}
    \cdots
    L'^{j_{f\dash}}_{\ta_{f-2} \ta_{f\dash}}
    C^{j_f}_{\ta_{f\dash}}
    \ket{\Phi_J}.\nonumber
  \end{align}
  Because the projector $\hat{\cal P}^{-}_p$ is absent for $p=f$ in
  Eq.(\ref{eq:tang_proj}), only Step \ref{step:expH} is executed
  at the rightmost site and now the wavefunction is expressed as
  \begin{align}
%    \ket{\Psi} =
    \sum_{{\bm \ta},J}
    L'^{j_1}_{\ta_1}
    L'^{j_2}_{\ta_1 \ta_{2}}
    \cdots
    L'^{j_{f\dash}}_{\ta_{f-2} \ta_{f\dash}}
    C'^{j_f}_{\ta_{f\dash}}
    \ket{\Phi_J}.\nonumber
  \end{align}
  This is the end of the former half propagation with a left-to-right
  sweep, then the latter half propagation will be done in the opposite
  direction, i.e.\ a right-to-left sweep, in the steps below.
\item The same as Step \ref{step:expH}; the coefficient of the current
  site $C'^{j_p}_{\ta_{p\dash} \ta_{p}}$ is propagated to
  $C''^{j_p}_{\ta_{p\dash} \ta_{p}}$ by applying the operator
  exponential $e^{-i\hat{\cal P}^+_p\hat{H}t/2}$.
  \label{step:expH_back}
\item The wavefunction after the propagation in the previous
  step is expressed as
  \begin{align}
    %\ket{\Psi^\al} =
    %\sum_{\ta_1\cdots\ta_{f\dash},J}
    \sum_{{\bm \ta},J}
    L'^{j_1}_{\ta_1}
    \cdots
    L'^{j_{p\dash}}_{\ta_{p\text{-2}} \ta_{p\dash}}
   (C''^{j_{p  }}_{\ta_{p\dash} \ta_{p  }})
    R''^{j_{p+1}}_{\ta_{p  } \ta_{p+1}}
    \cdots
    R''^{j_f}_{\ta_{f\dash}} \ket{{\Phi^{\al}_{J}}}.\nonumber
  \end{align}
  It is transformed to
  \begin{align}
    %\ket{\Psi^\al} =
    %\sum_{\ta_1\cdots\ta_{f\dash},J}
    \sum_{{\bm \ta},\ga_p,J}
    L'^{j_1}_{\ta_1}
    \cdots
    L'^{j_{p\dash}}_{\ta_{p\text{-2}} \ta_{p\dash}}
   (\si''^{\ta_{p\dash}}_{\ga_{p\dash}}
    R''^{j_{p  }}_{\ga_{p\dash} \ta_{p  }})
    R''^{j_{p+1}}_{\ta_{p  } \ta_{p+1}}
    \cdots
    R''^{j_f}_{\ta_{f\dash}} \ket{{\Phi^{\al}_{J}}}.\nonumber
  \end{align}
  by using the orthogonal decomposition in Eq.(\ref{eq:gauge_trf})
%  \begin{align}
%    C''^{j_{p }}_{\ta_{p\dash} \ta_{p}}
%    =
%    \sum_{\ga_{p}}
%    U_{\ta_{p\dash} \ga_{p\dash}}
%    \la_{\ga_{p\dash}}
%    R''^{j_{p  }}_{\ga_{p\dash} \ta_{p}}
%    =
%    \sum_{\ta'_{p}}
%    \si'^{\ta_{p\dash}}_{\ga_{p\dash}}
%    R''^{j_{p  }}_{\ga_{p\dash} \ta_{p}}.
%  \end{align}
%
\item The same as Step \ref{step:expK}; the matrix
  $\si''_{\ta_{p\dash}\ga_{p\dash}}$ is propagated backward in time to
  $\si'_{\ta_{p\dash}\ga_{p\dash}}$ by applying the operator
  exponential $e^{+i\hat{\cal P}^-_{p\dash}\hat{H}\de}$.
\item Similar to Step \ref{step:SR2C}, the wavefunction expressed as
  \begin{align}
    %\ket{\Psi^\al} =
    %\sum_{\ta_1\cdots\ta_{f\dash},J}
    \sum_{{\bm \ta},\ga_p,J}
    L'^{j_1}_{\ta_1}
    \cdots
   (L'^{j_{p\dash}}_{\ta_{p\text{-2}} \ga_{p\dash}}
    \si'^{\ga_{p\dash}}_{\ta_{p\dash}})
    R''^{j_{p  }}_{\ta_{p\dash} \ta_{p  }}
    \cdots
    R''^{j_f}_{\ta_{f\dash}} \ket{{\Phi^{\al}_{J}}},\nonumber
  \end{align}
  is transformed to
  \begin{align}
    %\ket{\Psi^\al} =
    %\sum_{\ta_1\cdots\ta_{f\dash},J}
    \sum_{{\bm \ta},J}
    L'^{j_1}_{\ta_1}
    \cdots
   (C'^{j_{p\dash}}_{\ta_{p\text{-2}} \ta_{p\dash}})
    R''^{j_{p  }}_{\ta_{p\dash} \ta_{p  }}
    \cdots
    R''^{j_f}_{\ta_{f\dash}} \ket{{\Phi^{\al}_{J}}}.\nonumber
  \end{align}
  \label{step:LS2C}
\item Step \ref{step:expH_back}--\ref{step:LS2C} are repeated until the
  current site $p$ reaches the leftmost site $p$=1
  \begin{align}
%    \Psi^{\al} =
    \sum_{{\bm \ta},J}
    C'^{j_1}_{\ta_1}
    R''^{j_2}_{\ta_1 \ta_{2}}
    \cdots
    R''^{j_{f\dash}}_{\ta_{f-2} \ta_{f\dash}}
    R''^{j_f}_{\ta_{f\dash}}
    \ket{\Phi_J}.\nonumber
  \end{align}
  Lastly, the coefficient $C'^{j_1}_{\ta_1}$ is propagated to
  $C''^{j_1}_{\ta_1}$ in the same way as Step \ref{step:expH_back}.
%  Now, we have the MPS site coefficients propagated by $2\de$ in time
%  \begin{align}
%%    \Psi^{\al} =
%    \sum_{{\bm \ta},J}
%    C''^{j_1}_{\ta_1}
%    R''^{j_2}_{\ta_1 \ta_{2}}
%    \cdots
%    R''^{j_{f\dash}}_{\ta_{f-2} \ta_{f\dash}}
%    R''^{j_f}_{\ta_{f\dash}}
%    \ket{\Phi_J}.\nonumber
%  \end{align}
\end{enumerate}
}

\section*{Appendix B: Comparison with the multilayer formulation}
A different form of the equation of motion (EOM) for the MPS-MCTDH
wavefunction ansatz defined in Eq.(\ref{eq:wf_mps_1}) with
Eq.(\ref{eq:wf_mps_2}) and (\ref{eq:wf_mps_3}) can be derived in the
framework of the multi-layer formulation,\cite{Wang:2003fu,
  Manthe:2008ev, Vendrell:2011fh} in which an immutable ML-tree
structure is defined, $e.g.$ if we choose the site coefficient at $p$,
namely $A^{j_p}_{\ta_{p\dash}\ta_p}$, as the top layer coefficient,
the definition of the ML-tree is given as \\[1ex]

{\it Top layer}
\begin{align}
\ket{\Psi}
&=A^{(1)j_p}_{\ta_{p\dash}\ta_p}
\ket{\Psi^{\Ltxx}_{\ta_{p\dash}} \vphi_{j_p} \Psi^{\Rtxx}_{\ta_{p}}},%\nonumber\\
\end{align}

{\it Left tree}
\begin{align}
\ket{\Psi^{\Ltxx}_{                                   \ta_{p\dash}}}
&=A^{(2)    j_{p\dash}}_{            \ta_{p\text{-2}}\ta_{p\dash}}
\ket{\Psi^{\Ltxx}_{\ta_{p\text{-2}}}  \vphi_{j_{p\dash}}},\nonumber\\
\ket{\Psi^{\Ltxx}_{                                   \ta_{p\text{-2}}}}
&=A^{(3)    j_{p\text{-2}}}_{            \ta_{p\text{-3}}\ta_{p\text{-2}}}
\ket{\Psi^{\Ltxx}_{\ta_{p\text{-3}}}  \vphi_{j_{p\text{-2}}}},\nonumber\\
&\;\:\vdots\nonumber\\
\ket{\Psi^{\Ltxx}_{\ta_{2}}}
&=A^{(p\dash)j_{2}}_{\ta_{1}\ta_{2}}
\ket{\Psi^\Ltxx_{\ta_{1}} \vphi_{j_{2}}},\nonumber\\
\ket{\Psi^\Ltxx_{\ta_{1}}}
&=A^{(p)j_{1}}_{\ta_{1}}\ket{\vphi_{j_{1}}},\nonumber
\end{align}

{\it Right tree}
\begin{align}
\ket{\Psi^{\Rtxx}_{                   \ta_{p     }}}
&=A^{(2)    j_{p+1    }}_{           \ta_{p     }\ta_{p+1}}
\ket{\vphi_{j_{p+1    }} \Psi^{\Rtxx}_{           \ta_{p+1}}},
\nonumber\\
\ket{\Psi^{\Rtxx}_{                   \ta_{p+1   }}}
&=A^{(3)    j_{p+2    }}_{           \ta_{p+1   }\ta_{p+2}}
\ket{\vphi_{j_{p+2    }} \Psi^{\Rtxx}_{           \ta_{p+2}}},
\nonumber\\
&\vdots\nonumber\\
\ket{\Psi^{\Rtxx}_{\ta_{f\text{-2}}}}
&=A^{(f\text{-}p)j_{f\dash}}_{\ta_{f\text{-2}}\ta_{f\dash}}
\ket{\vphi_{j_{f\dash}}\Psi^{\Rtxx}_{\ta_{f\dash}}},\nonumber\\
\ket{\Psi^\Rtxx_{\ta_{f\dash}}}
&=A^{(f\text{-}p+1)j_f}_{\ta_{f\dash}}\ket{\vphi_{j_{f}}},\nonumber
\end{align}
where the parenthesis in the super script denotes the depth in the
ML-tree structure. In the standard ML-MCTDH
notation,\cite{Wang:2018kp} $A^{(3) j_{p\text{-2}}}_{
  \ta_{p\text{-3}}\ta_{p\text{-2}}}$ is usually noted as $A^{\text{\tiny
    3;L}}_{ \ta_{p\text{-2}};\ta_{p\text{-3}}, j_{p\text{-2}}}$ where
L in the superscript denotes the route from the root to the node under
consideration, but there are only two routes (L or R) in the MPS
ansatz.
%\begin{align}
%\ket{\Psi}
%&=A^{(1)j_1}_{\ta_1}\ket{\vphi_{j_1} \Psi^{\Rtxx}_{\ta_1}},\nonumber\\
%\ket{\Psi^{\Rtxx}_{\ta_1}}
%&=A^{(2)j_2}_{\ta_1\ta_2}\ket{\vphi_{j_2}\Psi^{\Rtxx}_{\ta_2}},\nonumber\\
%\ket{\Psi^{\Rtxx}_{\ta_2}}
%&=A^{(3)j_3}_{\ta_2\ta_3}\ket{\vphi_{j_3}\Psi^{\Rtxx}_{\ta_3}},\nonumber\\
%&\vdots\nonumber\\
%\ket{\Psi^{\Rtxx}_{\ta_{f\text{-2}}}}
%&=A^{(f\dash)j_{f\dash}}_{\ta_{f\text{-2}}\ta_{f\dash}}
%\ket{\vphi_{j_{f\dash}}\Psi^{\Rtxx}_{\ta_{f\dash}}},\nonumber\\
%\ket{\Psi^\Rtxx_{\ta_{f\dash}}}
%&=A^{(f)j_f}_{\ta_{f\dash}}\ket{\vphi_{j_{f}}},\nonumber
%\end{align}
In the ML-formulation, the generalized SPFs $\Psi^{\Ltxx(\Rtxx)}$ at
every layer should be kept in orthonormal which is equivalent to fix
the representation of the wavefunction in the specific $p$-canonical
form of the MPS written as
\begin{align}
%  \ket{\Psi} =
  \sum_{{\bm \ta},J}
  L^{(p)j_1}_{\ta_1}
  \cdots
  L^{(2)j_{p-1}}_{\ta_{p-2} \ta_{p-1}}
  C^{(1)j_{p  }}_{\ta_{p-1} \ta_{p  }}
  R^{(2)j_{p+1}}_{\ta_{p  } \ta_{p+1}}
  \cdots
  R^{(f\text{-}p+1)j_f}_{\ta_{f-1}}
  \ket{\Phi_J},\nonumber
\end{align}
throughout the propagation, which is in contrast to the
MPS-formulation, in which the gauges of the site coefficients are
consecutively changed as Eq.(\ref{eq:gauge_trf}) and the
representation for the wavefunction is transformed between the
canonical forms of the different sites.
%$L^{j_{q }}_{\ta_{q-1} \ta_{q }}$ and $R^{j_{q
%}}_{\ta_{q-1} \ta_{q }}$
%
%so that the left- and right-renormalized
%basis in Eq.(\ref{eq:lbasis}) and (\ref{eq:rbasis}) always fulfill the
%orthonormal condition.
%
In the ML-formulation, because the representation of the wavefunction
is fixed as the canonical form of the specific site $p$, the basis
states for the generalized single hole functions (SHF) are
non-orthogonal except for the top layer and the inversion of the
overlap matrices of the SHFs, which can be singular sometimes, appear
in the EOMs for the lower layers, whereas in the MPS-formulation the
wavefunction is always expressed by the direct products of the
orthonormal basis states at every node of the tree and the inverse
matrices are completely eliminated in the EOMs.

The time propagation with the EOMs derived in the MPS-formulation is,
therefore, robust even for the systems with many layers. Such systems
can be found in an interesting application\cite{Schroder:2016jza} in
which infinite bath modes are mapped onto an effective 1-D chain
modes to efficiently simulate the open quantum dynamics beyond the
perturbation theory.\cite{Prior:2010gd, Plenio:2016dk}
%  M.H. Beck et al. Physics Reports 324 (2000) 1 105
%  5.2.4. Solving the CMF equations
%->In addition, it is known that the convergence of the DMRG XXXX is slow
%->as compared with the sweep algorithm, and thus the convergence to the
%->ground state by the imaginary time propagation should be accelerated
%->by adopting the EOMs of the MPS-formulation.
In addition, in contrast to the standard ML-formulation, the
MPS-formulation with the tangent space projector splitting method
allows us to adopt the Lanczos integrator, which possesses favorable
properties,\cite{Beck:2000wm} for the time propagation of
the site coefficients in spite of the highly non-linear
parametrization of the MPS ansatz,
because the differential equations obtained from the individual split
tangent space projectors are linear equations and exactly solvable.
The accurate time-integration algorithm should allow a comparatively
large step size for the propagation of the site coefficients, namely
the MCTDH expansion coefficients.

There is an ongoing argument regarding the pros and cons of the
projector-splitting integrator applied to the EOMs of the SPF
%, which is expanded by the time-independent primitive basis functions,
in the conventional MCTDH.\cite{Lubich:2015fr, Manthe:2015iq,
  Kloss:2017kq, Bonfanti:2018vj, Meyer:2018gr} While it can remove the
inversion of the density matrix completely from the EOMs and is robust
and not collapsed even when the density matrix has zero eigenvalues
caused by the presence of unoccupied natural orbitals, the propagation
of the unoccupied orbitals determined by the projector splitting
integrator will be somewhat arbitrary because it involves orthogonal
decompositions of the density matrix. This behavior is a consequence
of the nature of the first-order equation, and to be accurate, it is
not correct for the second-order in time as discussed in detail by
Manthe in Ref[\onlinecite{Manthe:2015iq}]. In the standard MCTDH
implementation, the problem of this inaccurate motion of the
unoccupied orbitals is usually mitigated by the regularization of the
density matrix,\cite{Beck:2000wm} which introduces an artificial
occupation to the unoccupied orbitals with a small number
$\varepsilon$ primarily to avoid the non-invertible density matrix
problem, since these orbitals occupied by the small number
$\varepsilon$are expected to be rotated quickly into their correct
direction due to the so-called {\it self-healing} effect of the
MCTDH. In fact, the recently developed new regularization
scheme,\cite{Meyer:2018gr, Wang:2018kp} which allows a much smaller
value for $\varepsilon$ comparing with the conventional scheme, has
exhibited an ability to rotate the unoccupied orbitals more quickly to
the correct directions.
The MPS-formulation in this paper also adopts the projector-splitting
integrator method and the situation seems to be much the same as
described above, for instance, if there is an unoccupied state
%in $\{\ket{\Psi^\Ltxx_{\ta_{p\dash}}\;\vphi_{j_p}}\}$
as $w^\text{\tiny (L)}_{\ta_p}=0$ in Eq.(\ref{eq:LwL}), the
transformation to the state $\Psi^\Ltxx_{\ta_{p+1}}$ becomes non-unique,
and thus further investigation is needed on that
point.%\cite{Bonfanti:2018vj}
%, $\ket{\Psi^\Ltxx_{\ta_p}} =
%L^{j_p}_{\ta_{p\dash}\ta_p} \ket{\Psi^\Ltxx_{\ta_{p\dash}}}$ in Step
%\ref{step:C2LS},
%, though the zero column of $\si'^{\ga_p}_{\ta_p}$ is
%soon filled by the propagation in \label{eq:expK}.
%

%general ML-tree XXX.
%
%\blindtext \cite{article-minimal}
\bibliographystyle{apsrev4-1} % Tell bibtex which bibliography style to use

\begin{thebibliography}{58}%
\makeatletter
\providecommand \@ifxundefined [1]{%
 \@ifx{#1\undefined}
}%
\providecommand \@ifnum [1]{%
 \ifnum #1\expandafter \@firstoftwo
 \else \expandafter \@secondoftwo
 \fi
}%
\providecommand \@ifx [1]{%
 \ifx #1\expandafter \@firstoftwo
 \else \expandafter \@secondoftwo
 \fi
}%
\providecommand \natexlab [1]{#1}%
\providecommand \enquote  [1]{``#1''}%
\providecommand \bibnamefont  [1]{#1}%
\providecommand \bibfnamefont [1]{#1}%
\providecommand \citenamefont [1]{#1}%
\providecommand \href@noop [0]{\@secondoftwo}%
\providecommand \href [0]{\begingroup \@sanitize@url \@href}%
\providecommand \@href[1]{\@@startlink{#1}\@@href}%
\providecommand \@@href[1]{\endgroup#1\@@endlink}%
\providecommand \@sanitize@url [0]{\catcode `\\12\catcode `\$12\catcode
  `\&12\catcode `\#12\catcode `\^12\catcode `\_12\catcode `\%12\relax}%
\providecommand \@@startlink[1]{}%
\providecommand \@@endlink[0]{}%
\providecommand \url  [0]{\begingroup\@sanitize@url \@url }%
\providecommand \@url [1]{\endgroup\@href {#1}{\urlprefix }}%
\providecommand \urlprefix  [0]{URL }%
\providecommand \Eprint [0]{\href }%
\providecommand \doibase [0]{http://dx.doi.org/}%
\providecommand \selectlanguage [0]{\@gobble}%
\providecommand \bibinfo  [0]{\@secondoftwo}%
\providecommand \bibfield  [0]{\@secondoftwo}%
\providecommand \translation [1]{[#1]}%
\providecommand \BibitemOpen [0]{}%
\providecommand \bibitemStop [0]{}%
\providecommand \bibitemNoStop [0]{.\EOS\space}%
\providecommand \EOS [0]{\spacefactor3000\relax}%
\providecommand \BibitemShut  [1]{\csname bibitem#1\endcsname}%
\let\auto@bib@innerbib\@empty
%</preamble>
\bibitem [{\citenamefont {White}(1992)}]{White:1992ie}%
  \BibitemOpen
  \bibfield  {author} {\bibinfo {author} {\bibfnamefont {S.~R.}\ \bibnamefont
  {White}},\ }\href@noop {} {\bibfield  {journal} {\bibinfo  {journal} {Phys.
  Rev. Lett.}\ }\textbf {\bibinfo {volume} {69}},\ \bibinfo {pages} {2863}
  (\bibinfo {year} {1992})}\BibitemShut {NoStop}%
\bibitem [{\citenamefont {White}(1993)}]{White:1993tn}%
  \BibitemOpen
  \bibfield  {author} {\bibinfo {author} {\bibfnamefont {S.~R.}\ \bibnamefont
  {White}},\ }\href@noop {} {\bibfield  {journal} {\bibinfo  {journal} {Phys.
  Rev. B}\ }\textbf {\bibinfo {volume} {48}},\ \bibinfo {pages} {10345}
  (\bibinfo {year} {1993})}\BibitemShut {NoStop}%
\bibitem [{\citenamefont {White}\ and\ \citenamefont
  {Martin}(1999)}]{White:1999ws}%
  \BibitemOpen
  \bibfield  {author} {\bibinfo {author} {\bibfnamefont {S.~R.}\ \bibnamefont
  {White}}\ and\ \bibinfo {author} {\bibfnamefont {R.~L.}\ \bibnamefont
  {Martin}},\ }\href@noop {} {\bibfield  {journal} {\bibinfo  {journal} {J.
  Chem. Phys.}\ }\textbf {\bibinfo {volume} {110}},\ \bibinfo {pages} {4127}
  (\bibinfo {year} {1999})}\BibitemShut {NoStop}%
\bibitem [{\citenamefont {Chan}\ and\ \citenamefont
  {Head-Gordon}(2002)}]{Chan:2002cz}%
  \BibitemOpen
  \bibfield  {author} {\bibinfo {author} {\bibfnamefont {G.~K.-L.}\
  \bibnamefont {Chan}}\ and\ \bibinfo {author} {\bibfnamefont {M.}~\bibnamefont
  {Head-Gordon}},\ }\href@noop {} {\bibfield  {journal} {\bibinfo  {journal}
  {J. Chem. Phys.}\ }\textbf {\bibinfo {volume} {116}},\ \bibinfo {pages}
  {4462} (\bibinfo {year} {2002})}\BibitemShut {NoStop}%
\bibitem [{\citenamefont {White}\ and\ \citenamefont
  {Feiguin}(2004)}]{White:2004fd}%
  \BibitemOpen
  \bibfield  {author} {\bibinfo {author} {\bibfnamefont {S.~R.}\ \bibnamefont
  {White}}\ and\ \bibinfo {author} {\bibfnamefont {A.~E.}\ \bibnamefont
  {Feiguin}},\ }\href@noop {} {\bibfield  {journal} {\bibinfo  {journal} {Phys.
  Rev. Lett.}\ }\textbf {\bibinfo {volume} {93}},\ \bibinfo {pages} {076401}
  (\bibinfo {year} {2004})}\BibitemShut {NoStop}%
\bibitem [{\citenamefont {Daley}\ \emph {et~al.}(2004)\citenamefont {Daley},
  \citenamefont {Kollath}, \citenamefont {Schollw{\"o}ck},\ and\ \citenamefont
  {Vidal}}]{Daley:2004hk}%
  \BibitemOpen
  \bibfield  {author} {\bibinfo {author} {\bibfnamefont {A.~J.}\ \bibnamefont
  {Daley}}, \bibinfo {author} {\bibfnamefont {C.}~\bibnamefont {Kollath}},
  \bibinfo {author} {\bibfnamefont {U.}~\bibnamefont {Schollw{\"o}ck}}, \ and\
  \bibinfo {author} {\bibfnamefont {G.}~\bibnamefont {Vidal}},\ }\href@noop {}
  {\bibfield  {journal} {\bibinfo  {journal} {J. Stat. Mech.}\ }\textbf
  {\bibinfo {volume} {2004}},\ \bibinfo {pages} {P04005} (\bibinfo {year}
  {2004})}\BibitemShut {NoStop}%
\bibitem [{\citenamefont {Vidal}(2004)}]{Vidal:2004jc}%
  \BibitemOpen
  \bibfield  {author} {\bibinfo {author} {\bibfnamefont {G.}~\bibnamefont
  {Vidal}},\ }\href@noop {} {\bibfield  {journal} {\bibinfo  {journal} {Phys.
  Rev. Lett.}\ }\textbf {\bibinfo {volume} {93}},\ \bibinfo {pages} {040502}
  (\bibinfo {year} {2004})}\BibitemShut {NoStop}%
\bibitem [{\citenamefont {Haegeman}\ \emph {et~al.}(2011)\citenamefont
  {Haegeman}, \citenamefont {Cirac}, \citenamefont {Osborne}, \citenamefont
  {Pi{\v z}orn}, \citenamefont {Verschelde},\ and\ \citenamefont
  {Verstraete}}]{Haegeman:2011dz}%
  \BibitemOpen
  \bibfield  {author} {\bibinfo {author} {\bibfnamefont {J.}~\bibnamefont
  {Haegeman}}, \bibinfo {author} {\bibfnamefont {J.~I.}\ \bibnamefont {Cirac}},
  \bibinfo {author} {\bibfnamefont {T.~J.}\ \bibnamefont {Osborne}}, \bibinfo
  {author} {\bibfnamefont {I.}~\bibnamefont {Pi{\v z}orn}}, \bibinfo {author}
  {\bibfnamefont {H.}~\bibnamefont {Verschelde}}, \ and\ \bibinfo {author}
  {\bibfnamefont {F.}~\bibnamefont {Verstraete}},\ }\href@noop {} {\bibfield
  {journal} {\bibinfo  {journal} {Phys. Rev. Lett.}\ }\textbf {\bibinfo
  {volume} {107}},\ \bibinfo {pages} {070601} (\bibinfo {year}
  {2011})}\BibitemShut {NoStop}%
\bibitem [{\citenamefont {Lubich}\ \emph {et~al.}(2013)\citenamefont {Lubich},
  \citenamefont {Rohwedder}, \citenamefont {Schneider},\ and\ \citenamefont
  {Vandereycken}}]{Lubich:2013ku}%
  \BibitemOpen
  \bibfield  {author} {\bibinfo {author} {\bibfnamefont {C.}~\bibnamefont
  {Lubich}}, \bibinfo {author} {\bibfnamefont {T.}~\bibnamefont {Rohwedder}},
  \bibinfo {author} {\bibfnamefont {R.}~\bibnamefont {Schneider}}, \ and\
  \bibinfo {author} {\bibfnamefont {B.}~\bibnamefont {Vandereycken}},\
  }\href@noop {} {\bibfield  {journal} {\bibinfo  {journal} {SIAM J. Matrix
  Anal. Appl.}\ }\textbf {\bibinfo {volume} {34}},\ \bibinfo {pages} {470}
  (\bibinfo {year} {2013})}\BibitemShut {NoStop}%
\bibitem [{\citenamefont {Haegeman}\ \emph {et~al.}(2013)\citenamefont
  {Haegeman}, \citenamefont {Osborne},\ and\ \citenamefont
  {Verstraete}}]{Haegeman:2013bg}%
  \BibitemOpen
  \bibfield  {author} {\bibinfo {author} {\bibfnamefont {J.}~\bibnamefont
  {Haegeman}}, \bibinfo {author} {\bibfnamefont {T.~J.}\ \bibnamefont
  {Osborne}}, \ and\ \bibinfo {author} {\bibfnamefont {F.}~\bibnamefont
  {Verstraete}},\ }\href@noop {} {\bibfield  {journal} {\bibinfo  {journal}
  {Phys. Rev. B}\ }\textbf {\bibinfo {volume} {88}},\ \bibinfo {pages} {075133}
  (\bibinfo {year} {2013})}\BibitemShut {NoStop}%
\bibitem [{\citenamefont {Ueda}\ \emph {et~al.}(2006)\citenamefont {Ueda},
  \citenamefont {Jin}, \citenamefont {Shibata}, \citenamefont {Hieida},\ and\
  \citenamefont {Nishino}}]{Ueda:2006tb}%
  \BibitemOpen
  \bibfield  {author} {\bibinfo {author} {\bibfnamefont {K.}~\bibnamefont
  {Ueda}}, \bibinfo {author} {\bibfnamefont {C.}~\bibnamefont {Jin}}, \bibinfo
  {author} {\bibfnamefont {N.}~\bibnamefont {Shibata}}, \bibinfo {author}
  {\bibfnamefont {Y.}~\bibnamefont {Hieida}}, \ and\ \bibinfo {author}
  {\bibfnamefont {T.}~\bibnamefont {Nishino}},\ }\href@noop {} {\bibfield
  {journal} {\bibinfo  {journal} {arXiv}\ } (\bibinfo {year} {2006})},\ \Eprint
  {http://arxiv.org/abs/cond-mat/0612480v2} {cond-mat/0612480v2} \BibitemShut
  {NoStop}%
\bibitem [{\citenamefont {Dorando}\ \emph {et~al.}(2009)\citenamefont
  {Dorando}, \citenamefont {Hachmann},\ and\ \citenamefont
  {Chan}}]{Dorando:2009dz}%
  \BibitemOpen
  \bibfield  {author} {\bibinfo {author} {\bibfnamefont {J.~J.}\ \bibnamefont
  {Dorando}}, \bibinfo {author} {\bibfnamefont {J.}~\bibnamefont {Hachmann}}, \
  and\ \bibinfo {author} {\bibfnamefont {G.~K.-L.}\ \bibnamefont {Chan}},\
  }\href@noop {} {\bibfield  {journal} {\bibinfo  {journal} {J. Chem. Phys.}\
  }\textbf {\bibinfo {volume} {130}},\ \bibinfo {pages} {184111} (\bibinfo
  {year} {2009})}\BibitemShut {NoStop}%
\bibitem [{\citenamefont {Kinder}\ \emph {et~al.}(2011)\citenamefont {Kinder},
  \citenamefont {Ralph},\ and\ \citenamefont {Chan}}]{Kinder:2011tq}%
  \BibitemOpen
  \bibfield  {author} {\bibinfo {author} {\bibfnamefont {J.~M.}\ \bibnamefont
  {Kinder}}, \bibinfo {author} {\bibfnamefont {C.~C.}\ \bibnamefont {Ralph}}, \
  and\ \bibinfo {author} {\bibfnamefont {G.~K.-L.}\ \bibnamefont {Chan}},\
  }\href@noop {} {\bibfield  {journal} {\bibinfo  {journal} {arXiv}\ }
  (\bibinfo {year} {2011})},\ \Eprint {http://arxiv.org/abs/1103.2155v1}
  {1103.2155v1} \BibitemShut {NoStop}%
\bibitem [{\citenamefont {Kinder}\ \emph {et~al.}(2014)\citenamefont {Kinder},
  \citenamefont {Ralph},\ and\ \citenamefont {Chan}}]{kinderquantum}%
  \BibitemOpen
  \bibfield  {author} {\bibinfo {author} {\bibfnamefont {J.}~\bibnamefont
  {Kinder}}, \bibinfo {author} {\bibfnamefont {C.}~\bibnamefont {Ralph}}, \
  and\ \bibinfo {author} {\bibfnamefont {G.}~\bibnamefont {Chan}},\ }\href@noop
  {} {\bibfield  {journal} {\bibinfo  {journal} {Advances in Chemical Physics}\
  }\textbf {\bibinfo {volume} {154}},\ \bibinfo {pages} {179} (\bibinfo {year}
  {2014})}\BibitemShut {NoStop}%
\bibitem [{\citenamefont {Wouters}\ \emph {et~al.}(2013)\citenamefont
  {Wouters}, \citenamefont {Nakatani}, \citenamefont {Van~Neck},\ and\
  \citenamefont {Chan}}]{wouters2013thouless}%
  \BibitemOpen
  \bibfield  {author} {\bibinfo {author} {\bibfnamefont {S.}~\bibnamefont
  {Wouters}}, \bibinfo {author} {\bibfnamefont {N.}~\bibnamefont {Nakatani}},
  \bibinfo {author} {\bibfnamefont {D.}~\bibnamefont {Van~Neck}}, \ and\
  \bibinfo {author} {\bibfnamefont {G.~K.-L.}\ \bibnamefont {Chan}},\
  }\href@noop {} {\bibfield  {journal} {\bibinfo  {journal} {Phys. Rev. B}\
  }\textbf {\bibinfo {volume} {88}},\ \bibinfo {pages} {075122} (\bibinfo
  {year} {2013})}\BibitemShut {NoStop}%
\bibitem [{\citenamefont {Nakatani}\ \emph {et~al.}(2014)\citenamefont
  {Nakatani}, \citenamefont {Wouters}, \citenamefont {Van~Neck},\ and\
  \citenamefont {Chan}}]{nakatani2014linear}%
  \BibitemOpen
  \bibfield  {author} {\bibinfo {author} {\bibfnamefont {N.}~\bibnamefont
  {Nakatani}}, \bibinfo {author} {\bibfnamefont {S.}~\bibnamefont {Wouters}},
  \bibinfo {author} {\bibfnamefont {D.}~\bibnamefont {Van~Neck}}, \ and\
  \bibinfo {author} {\bibfnamefont {G.~K.-L.}\ \bibnamefont {Chan}},\
  }\href@noop {} {\bibfield  {journal} {\bibinfo  {journal} {J. Chem. Phys.}\
  }\textbf {\bibinfo {volume} {140}},\ \bibinfo {pages} {024108} (\bibinfo
  {year} {2014})}\BibitemShut {NoStop}%
\bibitem [{\citenamefont {Haegeman}\ \emph {et~al.}(2016)\citenamefont
  {Haegeman}, \citenamefont {Lubich}, \citenamefont {Oseledets}, \citenamefont
  {Vandereycken},\ and\ \citenamefont {Verstraete}}]{Haegeman:2016fo}%
  \BibitemOpen
  \bibfield  {author} {\bibinfo {author} {\bibfnamefont {J.}~\bibnamefont
  {Haegeman}}, \bibinfo {author} {\bibfnamefont {C.}~\bibnamefont {Lubich}},
  \bibinfo {author} {\bibfnamefont {I.}~\bibnamefont {Oseledets}}, \bibinfo
  {author} {\bibfnamefont {B.}~\bibnamefont {Vandereycken}}, \ and\ \bibinfo
  {author} {\bibfnamefont {F.}~\bibnamefont {Verstraete}},\ }\href@noop {}
  {\bibfield  {journal} {\bibinfo  {journal} {Phys. Rev. B}\ }\textbf {\bibinfo
  {volume} {94}},\ \bibinfo {pages} {165116} (\bibinfo {year}
  {2016})}\BibitemShut {NoStop}%
\bibitem [{\citenamefont {Schr{\"o}der}\ and\ \citenamefont
  {Chin}(2016)}]{Schroder:2016jza}%
  \BibitemOpen
  \bibfield  {author} {\bibinfo {author} {\bibfnamefont {F.~A. Y.~N.}\
  \bibnamefont {Schr{\"o}der}}\ and\ \bibinfo {author} {\bibfnamefont {A.~W.}\
  \bibnamefont {Chin}},\ }\href@noop {} {\bibfield  {journal} {\bibinfo
  {journal} {Phys. Rev. B}\ }\textbf {\bibinfo {volume} {93}},\ \bibinfo
  {pages} {075105} (\bibinfo {year} {2016})}\BibitemShut {NoStop}%
\bibitem [{\citenamefont {Borrelli}\ and\ \citenamefont
  {Gelin}(2017)}]{Borrelli:2017jf}%
  \BibitemOpen
  \bibfield  {author} {\bibinfo {author} {\bibfnamefont {R.}~\bibnamefont
  {Borrelli}}\ and\ \bibinfo {author} {\bibfnamefont {M.~F.}\ \bibnamefont
  {Gelin}},\ }\href@noop {} {\bibfield  {journal} {\bibinfo  {journal} {Sci.
  Rep.}\ ,\ \bibinfo {pages} {1}} (\bibinfo {year} {2017})}\BibitemShut
  {NoStop}%
\bibitem [{\citenamefont {Kloss}\ \emph {et~al.}(2018)\citenamefont {Kloss},
  \citenamefont {Lev},\ and\ \citenamefont {Reichman}}]{Kloss:2018be}%
  \BibitemOpen
  \bibfield  {author} {\bibinfo {author} {\bibfnamefont {B.}~\bibnamefont
  {Kloss}}, \bibinfo {author} {\bibfnamefont {Y.~B.}\ \bibnamefont {Lev}}, \
  and\ \bibinfo {author} {\bibfnamefont {D.}~\bibnamefont {Reichman}},\
  }\href@noop {} {\bibfield  {journal} {\bibinfo  {journal} {Phys. Rev. B}\ ,\
  \bibinfo {pages} {1}} (\bibinfo {year} {2018})}\BibitemShut {NoStop}%
\bibitem [{\citenamefont {Meyer}\ \emph {et~al.}(1990)\citenamefont {Meyer},
  \citenamefont {Manthe},\ and\ \citenamefont {Cederbaum}}]{Meyer:1990if}%
  \BibitemOpen
  \bibfield  {author} {\bibinfo {author} {\bibfnamefont {H.-D.}\ \bibnamefont
  {Meyer}}, \bibinfo {author} {\bibfnamefont {U.}~\bibnamefont {Manthe}}, \
  and\ \bibinfo {author} {\bibfnamefont {L.~S.}\ \bibnamefont {Cederbaum}},\
  }\href@noop {} {\bibfield  {journal} {\bibinfo  {journal} {Chem. Phys.
  Lett.}\ }\textbf {\bibinfo {volume} {165}},\ \bibinfo {pages} {73} (\bibinfo
  {year} {1990})}\BibitemShut {NoStop}%
\bibitem [{\citenamefont {Manthe}\ \emph {et~al.}(1992)\citenamefont {Manthe},
  \citenamefont {Meyer},\ and\ \citenamefont {Cederbaum}}]{Manthe:1992cq}%
  \BibitemOpen
  \bibfield  {author} {\bibinfo {author} {\bibfnamefont {U.}~\bibnamefont
  {Manthe}}, \bibinfo {author} {\bibfnamefont {H.-D.}\ \bibnamefont {Meyer}}, \
  and\ \bibinfo {author} {\bibfnamefont {L.~S.}\ \bibnamefont {Cederbaum}},\
  }\href@noop {} {\bibfield  {journal} {\bibinfo  {journal} {J. Chem. Phys.}\
  }\textbf {\bibinfo {volume} {97}},\ \bibinfo {pages} {3199} (\bibinfo {year}
  {1992})}\BibitemShut {NoStop}%
\bibitem [{\citenamefont {Beck}\ \emph {et~al.}(2000)\citenamefont {Beck},
  \citenamefont {J{\"a}ckle}, \citenamefont {Worth},\ and\ \citenamefont
  {Meyer}}]{Beck:2000wm}%
  \BibitemOpen
  \bibfield  {author} {\bibinfo {author} {\bibfnamefont {M.~H.}\ \bibnamefont
  {Beck}}, \bibinfo {author} {\bibfnamefont {A.}~\bibnamefont {J{\"a}ckle}},
  \bibinfo {author} {\bibfnamefont {G.~A.}\ \bibnamefont {Worth}}, \ and\
  \bibinfo {author} {\bibfnamefont {H.-D.}\ \bibnamefont {Meyer}},\ }\href@noop
  {} {\bibfield  {journal} {\bibinfo  {journal} {Phys. Rep.}\ }\textbf
  {\bibinfo {volume} {324}},\ \bibinfo {pages} {1} (\bibinfo {year}
  {2000})}\BibitemShut {NoStop}%
\bibitem [{\citenamefont {Worth}\ \emph {et~al.}(1996)\citenamefont {Worth},
  \citenamefont {Meyer},\ and\ \citenamefont {Cederbaum}}]{Worth:1996ck}%
  \BibitemOpen
  \bibfield  {author} {\bibinfo {author} {\bibfnamefont {G.~A.}\ \bibnamefont
  {Worth}}, \bibinfo {author} {\bibfnamefont {H.-D.}\ \bibnamefont {Meyer}}, \
  and\ \bibinfo {author} {\bibfnamefont {L.~S.}\ \bibnamefont {Cederbaum}},\
  }\href@noop {} {\bibfield  {journal} {\bibinfo  {journal} {J. Chem. Phys.}\
  }\textbf {\bibinfo {volume} {105}},\ \bibinfo {pages} {4412} (\bibinfo {year}
  {1996})}\BibitemShut {NoStop}%
\bibitem [{\citenamefont {Raab}\ \emph {et~al.}(1999)\citenamefont {Raab},
  \citenamefont {Worth}, \citenamefont {Meyer},\ and\ \citenamefont
  {Cederbaum}}]{Raab:1999fa}%
  \BibitemOpen
  \bibfield  {author} {\bibinfo {author} {\bibfnamefont {A.}~\bibnamefont
  {Raab}}, \bibinfo {author} {\bibfnamefont {G.~A.}\ \bibnamefont {Worth}},
  \bibinfo {author} {\bibfnamefont {H.-D.}\ \bibnamefont {Meyer}}, \ and\
  \bibinfo {author} {\bibfnamefont {L.~S.}\ \bibnamefont {Cederbaum}},\
  }\href@noop {} {\bibfield  {journal} {\bibinfo  {journal} {J. Chem. Phys.}\
  }\textbf {\bibinfo {volume} {110}},\ \bibinfo {pages} {936} (\bibinfo {year}
  {1999})}\BibitemShut {NoStop}%
\bibitem [{\citenamefont {Grasedyck}(2010)}]{Grasedyck:2010cs}%
  \BibitemOpen
  \bibfield  {author} {\bibinfo {author} {\bibfnamefont {L.}~\bibnamefont
  {Grasedyck}},\ }\href@noop {} {\bibfield  {journal} {\bibinfo  {journal}
  {SIAM J. Matrix Anal. Appl.}\ }\textbf {\bibinfo {volume} {31}},\ \bibinfo
  {pages} {2029} (\bibinfo {year} {2010})}\BibitemShut {NoStop}%
\bibitem [{\citenamefont {Manthe}(2008)}]{Manthe:2008ev}%
  \BibitemOpen
  \bibfield  {author} {\bibinfo {author} {\bibfnamefont {U.}~\bibnamefont
  {Manthe}},\ }\href@noop {} {\bibfield  {journal} {\bibinfo  {journal} {J.
  Chem. Phys.}\ }\textbf {\bibinfo {volume} {128}},\ \bibinfo {pages} {164116}
  (\bibinfo {year} {2008})}\BibitemShut {NoStop}%
\bibitem [{\citenamefont {Wang}(2015)}]{Wang:2015ef}%
  \BibitemOpen
  \bibfield  {author} {\bibinfo {author} {\bibfnamefont {H.}~\bibnamefont
  {Wang}},\ }\href@noop {} {\bibfield  {journal} {\bibinfo  {journal} {J. Phys.
  Chem. A}\ }\textbf {\bibinfo {volume} {119}},\ \bibinfo {pages} {7951}
  (\bibinfo {year} {2015})}\BibitemShut {NoStop}%
\bibitem [{\citenamefont {Wang}\ and\ \citenamefont
  {Thoss}(2003)}]{Wang:2003fu}%
  \BibitemOpen
  \bibfield  {author} {\bibinfo {author} {\bibfnamefont {H.}~\bibnamefont
  {Wang}}\ and\ \bibinfo {author} {\bibfnamefont {M.}~\bibnamefont {Thoss}},\
  }\href@noop {} {\bibfield  {journal} {\bibinfo  {journal} {J. Chem. Phys.}\
  }\textbf {\bibinfo {volume} {119}},\ \bibinfo {pages} {1289} (\bibinfo {year}
  {2003})}\BibitemShut {NoStop}%
\bibitem [{\citenamefont {Vendrell}\ and\ \citenamefont
  {Meyer}(2011)}]{Vendrell:2011fh}%
  \BibitemOpen
  \bibfield  {author} {\bibinfo {author} {\bibfnamefont {O.}~\bibnamefont
  {Vendrell}}\ and\ \bibinfo {author} {\bibfnamefont {H.-D.}\ \bibnamefont
  {Meyer}},\ }\href@noop {} {\bibfield  {journal} {\bibinfo  {journal} {J.
  Chem. Phys.}\ }\textbf {\bibinfo {volume} {134}},\ \bibinfo {pages} {044135}
  (\bibinfo {year} {2011})}\BibitemShut {NoStop}%
\bibitem [{\citenamefont {Wang}\ and\ \citenamefont
  {Thoss}(2007)}]{Wang:2007eu}%
  \BibitemOpen
  \bibfield  {author} {\bibinfo {author} {\bibfnamefont {H.}~\bibnamefont
  {Wang}}\ and\ \bibinfo {author} {\bibfnamefont {M.}~\bibnamefont {Thoss}},\
  }\href@noop {} {\bibfield  {journal} {\bibinfo  {journal} {J. Phys. Chem. A}\
  }\textbf {\bibinfo {volume} {111}},\ \bibinfo {pages} {10369} (\bibinfo
  {year} {2007})}\BibitemShut {NoStop}%
\bibitem [{\citenamefont {Cao}\ \emph {et~al.}(2013)\citenamefont {Cao},
  \citenamefont {Kr{\"o}nke}, \citenamefont {Vendrell},\ and\ \citenamefont
  {Schmelcher}}]{Cao:2013kx}%
  \BibitemOpen
  \bibfield  {author} {\bibinfo {author} {\bibfnamefont {L.}~\bibnamefont
  {Cao}}, \bibinfo {author} {\bibfnamefont {S.}~\bibnamefont {Kr{\"o}nke}},
  \bibinfo {author} {\bibfnamefont {O.}~\bibnamefont {Vendrell}}, \ and\
  \bibinfo {author} {\bibfnamefont {P.}~\bibnamefont {Schmelcher}},\
  }\href@noop {} {\bibfield  {journal} {\bibinfo  {journal} {J. Chem. Phys.}\
  }\textbf {\bibinfo {volume} {139}},\ \bibinfo {pages} {134103} (\bibinfo
  {year} {2013})}\BibitemShut {NoStop}%
\bibitem [{\citenamefont {Manthe}(2017)}]{Manthe:2017hj}%
  \BibitemOpen
  \bibfield  {author} {\bibinfo {author} {\bibfnamefont {U.}~\bibnamefont
  {Manthe}},\ }\href@noop {} {\bibfield  {journal} {\bibinfo  {journal} {J.
  Phys.: Condens. Matter}\ }\textbf {\bibinfo {volume} {29}},\ \bibinfo {pages}
  {253001} (\bibinfo {year} {2017})}\BibitemShut {NoStop}%
\bibitem [{\citenamefont {Manthe}\ and\ \citenamefont
  {Weike}(2017)}]{Manthe:2017dx}%
  \BibitemOpen
  \bibfield  {author} {\bibinfo {author} {\bibfnamefont {U.}~\bibnamefont
  {Manthe}}\ and\ \bibinfo {author} {\bibfnamefont {T.}~\bibnamefont {Weike}},\
  }\href@noop {} {\bibfield  {journal} {\bibinfo  {journal} {J. Chem. Phys.}\
  }\textbf {\bibinfo {volume} {146}},\ \bibinfo {pages} {064117} (\bibinfo
  {year} {2017})}\BibitemShut {NoStop}%
\bibitem [{\citenamefont {K{\"u}hn}\ and\ \citenamefont
  {Lochbrunner}(2011)}]{Kuhn:2011tv}%
  \BibitemOpen
  \bibfield  {author} {\bibinfo {author} {\bibfnamefont {O.}~\bibnamefont
  {K{\"u}hn}}\ and\ \bibinfo {author} {\bibfnamefont {S.}~\bibnamefont
  {Lochbrunner}},\ }\href@noop {} {\bibfield  {journal} {\bibinfo  {journal}
  {arXiv}\ } (\bibinfo {year} {2011})},\ \Eprint
  {http://arxiv.org/abs/1108.4834v2} {1108.4834v2} \BibitemShut {NoStop}%
\bibitem [{\citenamefont {Schr{\"o}ter}\ \emph {et~al.}(2015)\citenamefont
  {Schr{\"o}ter}, \citenamefont {Ivanov}, \citenamefont {Schulze},
  \citenamefont {Polyutov}, \citenamefont {Yan}, \citenamefont {Pullerits},\
  and\ \citenamefont {Kuhn}}]{Schroter:2015je}%
  \BibitemOpen
  \bibfield  {author} {\bibinfo {author} {\bibfnamefont {M.}~\bibnamefont
  {Schr{\"o}ter}}, \bibinfo {author} {\bibfnamefont {S.~D.}\ \bibnamefont
  {Ivanov}}, \bibinfo {author} {\bibfnamefont {J.}~\bibnamefont {Schulze}},
  \bibinfo {author} {\bibfnamefont {S.~P.}\ \bibnamefont {Polyutov}}, \bibinfo
  {author} {\bibfnamefont {Y.}~\bibnamefont {Yan}}, \bibinfo {author}
  {\bibfnamefont {T.}~\bibnamefont {Pullerits}}, \ and\ \bibinfo {author}
  {\bibfnamefont {O.}~\bibnamefont {Kuhn}},\ }\href@noop {} {\bibfield
  {journal} {\bibinfo  {journal} {Phys. Rep.}\ }\textbf {\bibinfo {volume}
  {567}},\ \bibinfo {pages} {1} (\bibinfo {year} {2015})}\BibitemShut {NoStop}%
\bibitem [{\citenamefont {Schulze}\ \emph {et~al.}(2016)\citenamefont
  {Schulze}, \citenamefont {Shibl}, \citenamefont {Al-Marri},\ and\
  \citenamefont {K{\"u}hn}}]{Schulze:2016it}%
  \BibitemOpen
  \bibfield  {author} {\bibinfo {author} {\bibfnamefont {J.}~\bibnamefont
  {Schulze}}, \bibinfo {author} {\bibfnamefont {M.~F.}\ \bibnamefont {Shibl}},
  \bibinfo {author} {\bibfnamefont {M.~J.}\ \bibnamefont {Al-Marri}}, \ and\
  \bibinfo {author} {\bibfnamefont {O.}~\bibnamefont {K{\"u}hn}},\ }\href@noop
  {} {\bibfield  {journal} {\bibinfo  {journal} {J. Chem. Phys.}\ }\textbf
  {\bibinfo {volume} {144}},\ \bibinfo {pages} {185101} (\bibinfo {year}
  {2016})}\BibitemShut {NoStop}%
\bibitem [{\citenamefont {Shibl}\ \emph {et~al.}(2017)\citenamefont {Shibl},
  \citenamefont {Schulze}, \citenamefont {Al-Marri},\ and\ \citenamefont
  {K{\"u}hn}}]{Shibl:2017jl}%
  \BibitemOpen
  \bibfield  {author} {\bibinfo {author} {\bibfnamefont {M.~F.}\ \bibnamefont
  {Shibl}}, \bibinfo {author} {\bibfnamefont {J.}~\bibnamefont {Schulze}},
  \bibinfo {author} {\bibfnamefont {M.~J.}\ \bibnamefont {Al-Marri}}, \ and\
  \bibinfo {author} {\bibfnamefont {O.}~\bibnamefont {K{\"u}hn}},\ }\href@noop
  {} {\bibfield  {journal} {\bibinfo  {journal} {J. Phys. B: At. Mol. Opt.
  Phys.}\ }\textbf {\bibinfo {volume} {50}},\ \bibinfo {pages} {184001}
  (\bibinfo {year} {2017})}\BibitemShut {NoStop}%
\bibitem [{\citenamefont {Shi}\ \emph {et~al.}(2006)\citenamefont {Shi},
  \citenamefont {Duan},\ and\ \citenamefont {Vidal}}]{Shi:2006hz}%
  \BibitemOpen
  \bibfield  {author} {\bibinfo {author} {\bibfnamefont {Y.~Y.}\ \bibnamefont
  {Shi}}, \bibinfo {author} {\bibfnamefont {L.~M.}\ \bibnamefont {Duan}}, \
  and\ \bibinfo {author} {\bibfnamefont {G.}~\bibnamefont {Vidal}},\
  }\href@noop {} {\bibfield  {journal} {\bibinfo  {journal} {Phys. Rev. A}\
  }\textbf {\bibinfo {volume} {74}},\ \bibinfo {pages} {134} (\bibinfo {year}
  {2006})}\BibitemShut {NoStop}%
\bibitem [{\citenamefont {{Schollw{\"o}ck}}(2011)}]{Schollwoeck:2010gl}%
  \BibitemOpen
  \bibfield  {author} {\bibinfo {author} {\bibfnamefont {U.}~\bibnamefont
  {{Schollw{\"o}ck}}},\ }\href@noop {} {\bibfield  {journal} {\bibinfo
  {journal} {Ann. Phys.}\ }\textbf {\bibinfo {volume} {326}},\ \bibinfo {pages}
  {96} (\bibinfo {year} {2011})}\BibitemShut {NoStop}%
\bibitem [{\citenamefont {Lubich}(2015)}]{Lubich:2015fr}%
  \BibitemOpen
  \bibfield  {author} {\bibinfo {author} {\bibfnamefont {C.}~\bibnamefont
  {Lubich}},\ }\href@noop {} {\bibfield  {journal} {\bibinfo  {journal} {Appl.
  Math. Res. Express}\ }\textbf {\bibinfo {volume} {2015}},\ \bibinfo {pages}
  {311} (\bibinfo {year} {2015})}\BibitemShut {NoStop}%
\bibitem [{\citenamefont {Kloss}\ \emph {et~al.}(2017)\citenamefont {Kloss},
  \citenamefont {Burghardt},\ and\ \citenamefont {Lubich}}]{Kloss:2017kq}%
  \BibitemOpen
  \bibfield  {author} {\bibinfo {author} {\bibfnamefont {B.}~\bibnamefont
  {Kloss}}, \bibinfo {author} {\bibfnamefont {I.}~\bibnamefont {Burghardt}}, \
  and\ \bibinfo {author} {\bibfnamefont {C.}~\bibnamefont {Lubich}},\
  }\href@noop {} {\bibfield  {journal} {\bibinfo  {journal} {J. Chem. Phys.}\
  }\textbf {\bibinfo {volume} {146}},\ \bibinfo {pages} {174107} (\bibinfo
  {year} {2017})}\BibitemShut {NoStop}%
\bibitem [{\citenamefont {{Bonfanti}}\ and\ \citenamefont
  {{Burghardt}}(2018)}]{Bonfanti:2018vj}%
  \BibitemOpen
  \bibfield  {author} {\bibinfo {author} {\bibfnamefont {M.}~\bibnamefont
  {{Bonfanti}}}\ and\ \bibinfo {author} {\bibfnamefont {I.}~\bibnamefont
  {{Burghardt}}},\ }\href@noop {} {\bibfield  {journal} {\bibinfo  {journal}
  {ArXiv e-prints}\ } (\bibinfo {year} {2018})},\ \Eprint
  {http://arxiv.org/abs/1802.01058} {arXiv:1802.01058 [physics.chem-ph]}
  \BibitemShut {NoStop}%
\bibitem [{\citenamefont {Beck}\ and\ \citenamefont
  {Meyer}(1997)}]{Beck:1997fs}%
  \BibitemOpen
  \bibfield  {author} {\bibinfo {author} {\bibfnamefont {M.~H.}\ \bibnamefont
  {Beck}}\ and\ \bibinfo {author} {\bibfnamefont {H.-D.}\ \bibnamefont
  {Meyer}},\ }\href@noop {} {\bibfield  {journal} {\bibinfo  {journal} {Z.
  Phys. D}\ }\textbf {\bibinfo {volume} {42}},\ \bibinfo {pages} {113}
  (\bibinfo {year} {1997})}\BibitemShut {NoStop}%
\bibitem [{\citenamefont {Carter}\ \emph {et~al.}(1997)\citenamefont {Carter},
  \citenamefont {Culik},\ and\ \citenamefont {Bowman}}]{Carter:1997ej}%
  \BibitemOpen
  \bibfield  {author} {\bibinfo {author} {\bibfnamefont {S.}~\bibnamefont
  {Carter}}, \bibinfo {author} {\bibfnamefont {S.~J.}\ \bibnamefont {Culik}}, \
  and\ \bibinfo {author} {\bibfnamefont {J.~M.}\ \bibnamefont {Bowman}},\
  }\href@noop {} {\bibfield  {journal} {\bibinfo  {journal} {J. Chem. Phys.}\
  }\textbf {\bibinfo {volume} {107}},\ \bibinfo {pages} {10458} (\bibinfo
  {year} {1997})}\BibitemShut {NoStop}%
\bibitem [{\citenamefont {Yagi}\ \emph {et~al.}(2004)\citenamefont {Yagi},
  \citenamefont {Hirao}, \citenamefont {Taketsugu}, \citenamefont {Schmidt},\
  and\ \citenamefont {Gordon}}]{Yagi:2004jy}%
  \BibitemOpen
  \bibfield  {author} {\bibinfo {author} {\bibfnamefont {K.}~\bibnamefont
  {Yagi}}, \bibinfo {author} {\bibfnamefont {K.}~\bibnamefont {Hirao}},
  \bibinfo {author} {\bibfnamefont {T.}~\bibnamefont {Taketsugu}}, \bibinfo
  {author} {\bibfnamefont {M.~W.}\ \bibnamefont {Schmidt}}, \ and\ \bibinfo
  {author} {\bibfnamefont {M.~S.}\ \bibnamefont {Gordon}},\ }\href@noop {}
  {\bibfield  {journal} {\bibinfo  {journal} {J. Chem. Phys.}\ }\textbf
  {\bibinfo {volume} {121}},\ \bibinfo {pages} {1383} (\bibinfo {year}
  {2004})}\BibitemShut {NoStop}%
\bibitem [{\citenamefont {Kurashige}\ and\ \citenamefont
  {Yanai}(2011)}]{Kurashige:2011ck}%
  \BibitemOpen
  \bibfield  {author} {\bibinfo {author} {\bibfnamefont {Y.}~\bibnamefont
  {Kurashige}}\ and\ \bibinfo {author} {\bibfnamefont {T.}~\bibnamefont
  {Yanai}},\ }\href@noop {} {\bibfield  {journal} {\bibinfo  {journal} {J.
  Chem. Phys.}\ }\textbf {\bibinfo {volume} {135}},\ \bibinfo {pages} {094104}
  (\bibinfo {year} {2011})}\BibitemShut {NoStop}%
\bibitem [{\citenamefont {Tamura}\ \emph {et~al.}(2007)\citenamefont {Tamura},
  \citenamefont {Bittner},\ and\ \citenamefont {Burghardt}}]{Tamura:2007ke}%
  \BibitemOpen
  \bibfield  {author} {\bibinfo {author} {\bibfnamefont {H.}~\bibnamefont
  {Tamura}}, \bibinfo {author} {\bibfnamefont {E.~R.}\ \bibnamefont {Bittner}},
  \ and\ \bibinfo {author} {\bibfnamefont {I.}~\bibnamefont {Burghardt}},\
  }\href@noop {} {\bibfield  {journal} {\bibinfo  {journal} {J. Chem. Phys.}\
  }\textbf {\bibinfo {volume} {126}},\ \bibinfo {pages} {021103} (\bibinfo
  {year} {2007})}\BibitemShut {NoStop}%
\bibitem [{\citenamefont {Tamura}\ \emph {et~al.}(2012)\citenamefont {Tamura},
  \citenamefont {Martinazzo}, \citenamefont {Ruckenbauer},\ and\ \citenamefont
  {Burghardt}}]{Tamura:2012fv}%
  \BibitemOpen
  \bibfield  {author} {\bibinfo {author} {\bibfnamefont {H.}~\bibnamefont
  {Tamura}}, \bibinfo {author} {\bibfnamefont {R.}~\bibnamefont {Martinazzo}},
  \bibinfo {author} {\bibfnamefont {M.}~\bibnamefont {Ruckenbauer}}, \ and\
  \bibinfo {author} {\bibfnamefont {I.}~\bibnamefont {Burghardt}},\ }\href@noop
  {} {\bibfield  {journal} {\bibinfo  {journal} {J. Chem. Phys.}\ }\textbf
  {\bibinfo {volume} {137}},\ \bibinfo {pages} {22A540} (\bibinfo {year}
  {2012})}\BibitemShut {NoStop}%
\bibitem [{\citenamefont {Tamura}\ \emph {et~al.}(2015)\citenamefont {Tamura},
  \citenamefont {Huix-Rotllant}, \citenamefont {Burghardt}, \citenamefont
  {Olivier},\ and\ \citenamefont {Beljonne}}]{Tamura:2015il}%
  \BibitemOpen
  \bibfield  {author} {\bibinfo {author} {\bibfnamefont {H.}~\bibnamefont
  {Tamura}}, \bibinfo {author} {\bibfnamefont {M.}~\bibnamefont
  {Huix-Rotllant}}, \bibinfo {author} {\bibfnamefont {I.}~\bibnamefont
  {Burghardt}}, \bibinfo {author} {\bibfnamefont {Y.}~\bibnamefont {Olivier}},
  \ and\ \bibinfo {author} {\bibfnamefont {D.}~\bibnamefont {Beljonne}},\
  }\href@noop {} {\bibfield  {journal} {\bibinfo  {journal} {Phys. Rev. Lett.}\
  }\textbf {\bibinfo {volume} {115}},\ \bibinfo {pages} {107401} (\bibinfo
  {year} {2015})}\BibitemShut {NoStop}%
\bibitem [{\citenamefont {Binder}\ \emph {et~al.}(2017)\citenamefont {Binder},
  \citenamefont {Polkehn}, \citenamefont {Ma},\ and\ \citenamefont
  {Burghardt}}]{Binder:2017hv}%
  \BibitemOpen
  \bibfield  {author} {\bibinfo {author} {\bibfnamefont {R.}~\bibnamefont
  {Binder}}, \bibinfo {author} {\bibfnamefont {M.}~\bibnamefont {Polkehn}},
  \bibinfo {author} {\bibfnamefont {T.}~\bibnamefont {Ma}}, \ and\ \bibinfo
  {author} {\bibfnamefont {I.}~\bibnamefont {Burghardt}},\ }\href@noop {}
  {\bibfield  {journal} {\bibinfo  {journal} {Chem. Phys.}\ }\textbf {\bibinfo
  {volume} {482}},\ \bibinfo {pages} {16} (\bibinfo {year} {2017})}\BibitemShut
  {NoStop}%
\bibitem [{\citenamefont {Ren}\ \emph {et~al.}(2018)\citenamefont {Ren},
  \citenamefont {Shuai},\ and\ \citenamefont {Chan}}]{Ren:2018tf}%
  \BibitemOpen
  \bibfield  {author} {\bibinfo {author} {\bibfnamefont {J.}~\bibnamefont
  {Ren}}, \bibinfo {author} {\bibfnamefont {Z.}~\bibnamefont {Shuai}}, \ and\
  \bibinfo {author} {\bibfnamefont {G.~K.-L.}\ \bibnamefont {Chan}},\
  }\href@noop {} {\bibfield  {journal} {\bibinfo  {journal} {arXiv}\ }
  (\bibinfo {year} {2018})},\ \Eprint {http://arxiv.org/abs/1806.07443}
  {1806.07443} \BibitemShut {NoStop}%
\bibitem [{\citenamefont {Kurashige}\ and\ \citenamefont
  {Yanai}(2009)}]{Kurashige:2009gs}%
  \BibitemOpen
  \bibfield  {author} {\bibinfo {author} {\bibfnamefont {Y.}~\bibnamefont
  {Kurashige}}\ and\ \bibinfo {author} {\bibfnamefont {T.}~\bibnamefont
  {Yanai}},\ }\href@noop {} {\bibfield  {journal} {\bibinfo  {journal} {J.
  Chem. Phys.}\ }\textbf {\bibinfo {volume} {130}},\ \bibinfo {pages} {234114}
  (\bibinfo {year} {2009})}\BibitemShut {NoStop}%
\bibitem [{\citenamefont {Prior}\ \emph {et~al.}(2010)\citenamefont {Prior},
  \citenamefont {Chin}, \citenamefont {Huelga},\ and\ \citenamefont
  {Plenio}}]{Prior:2010gd}%
  \BibitemOpen
  \bibfield  {author} {\bibinfo {author} {\bibfnamefont {J.}~\bibnamefont
  {Prior}}, \bibinfo {author} {\bibfnamefont {A.~W.}\ \bibnamefont {Chin}},
  \bibinfo {author} {\bibfnamefont {S.~F.}\ \bibnamefont {Huelga}}, \ and\
  \bibinfo {author} {\bibfnamefont {M.~B.}\ \bibnamefont {Plenio}},\
  }\href@noop {} {\bibfield  {journal} {\bibinfo  {journal} {Phys. Rev. Lett.}\
  }\textbf {\bibinfo {volume} {105}},\ \bibinfo {pages} {050404} (\bibinfo
  {year} {2010})}\BibitemShut {NoStop}%
\bibitem [{\citenamefont {Rosenbach}\ \emph {et~al.}(2016)\citenamefont
  {Rosenbach}, \citenamefont {Cerrillo}, \citenamefont {Huelga}, \citenamefont
  {Cao},\ and\ \citenamefont {Plenio}}]{Plenio:2016dk}%
  \BibitemOpen
  \bibfield  {author} {\bibinfo {author} {\bibfnamefont {R.}~\bibnamefont
  {Rosenbach}}, \bibinfo {author} {\bibfnamefont {J.}~\bibnamefont {Cerrillo}},
  \bibinfo {author} {\bibfnamefont {S.~F.}\ \bibnamefont {Huelga}}, \bibinfo
  {author} {\bibfnamefont {J.}~\bibnamefont {Cao}}, \ and\ \bibinfo {author}
  {\bibfnamefont {M.~B.}\ \bibnamefont {Plenio}},\ }\href@noop {} {\bibfield
  {journal} {\bibinfo  {journal} {New J. Phys.}\ }\textbf {\bibinfo {volume}
  {18}},\ \bibinfo {pages} {023035} (\bibinfo {year} {2016})}\BibitemShut
  {NoStop}%
\bibitem [{\citenamefont {Manthe}(2015)}]{Manthe:2015iq}%
  \BibitemOpen
  \bibfield  {author} {\bibinfo {author} {\bibfnamefont {U.}~\bibnamefont
  {Manthe}},\ }\href@noop {} {\bibfield  {journal} {\bibinfo  {journal} {J.
  Chem. Phys.}\ }\textbf {\bibinfo {volume} {142}},\ \bibinfo {pages} {244109}
  (\bibinfo {year} {2015})}\BibitemShut {NoStop}%
\bibitem [{\citenamefont {Meyer}\ and\ \citenamefont
  {Wang}(2018)}]{Meyer:2018gr}%
  \BibitemOpen
  \bibfield  {author} {\bibinfo {author} {\bibfnamefont {H.-D.}\ \bibnamefont
  {Meyer}}\ and\ \bibinfo {author} {\bibfnamefont {H.}~\bibnamefont {Wang}},\
  }\href@noop {} {\bibfield  {journal} {\bibinfo  {journal} {J. Chem. Phys.}\
  }\textbf {\bibinfo {volume} {148}},\ \bibinfo {pages} {124105} (\bibinfo
  {year} {2018})}\BibitemShut {NoStop}%
\bibitem [{\citenamefont {Wang}\ and\ \citenamefont
  {Meyer}(2018)}]{Wang:2018kp}%
  \BibitemOpen
  \bibfield  {author} {\bibinfo {author} {\bibfnamefont {H.}~\bibnamefont
  {Wang}}\ and\ \bibinfo {author} {\bibfnamefont {H.-D.}\ \bibnamefont
  {Meyer}},\ }\href@noop {} {\bibfield  {journal} {\bibinfo  {journal} {J.
  Chem. Phys.}\ }\textbf {\bibinfo {volume} {149}},\ \bibinfo {pages} {044119}
  (\bibinfo {year} {2018})}\BibitemShut {NoStop}%
\end{thebibliography}

\end{document}